\documentclass{aa}

 \pdfoutput=1
 
 \usepackage{natbib}
 \usepackage{txfonts}
 \usepackage{gensymb}
 \usepackage{inputenc}
 \usepackage{float}
 \usepackage{hyperref}
 \usepackage{amsmath}
 \usepackage{soul}
 
 \bibpunct{(}{)}{;}{a}{}{,}

 \usepackage{color}
 \usepackage{url}
 \usepackage{caption}

 \usepackage{breakurl}
 \usepackage{graphicx}

 \usepackage{multirow}
 \usepackage{subfig}  
 
 \defcitealias{CF00}{CF00}
 
\usepackage{color}

\begin{document}
        
        \title{HELP: modelling the spectral energy distributions of  \textit{Herschel} detected galaxies in the ELAIS~N1 field}
        \author{K.~Małek\inst{1,2}\thanks{\email{Katarzyna.Malek@ncbj.gov.pl}} \and V.~Buat\inst{1} \and Y.~Roehlly\inst{3,4} \and D.~Burgarella\inst{1}  \and  P.~D.~Hurley\inst{3}  \and R.~Shirley\inst{3} \and K.~Duncan\inst{5} \and A.~Efstathiou\inst{6} \and  A.~Papadopoulos\inst{6} \and M.~Vaccari\inst{7,8} \and D.~Farrah\inst{9} \and L.~Marchetti\inst{8,10} \and S.~Oliver\inst{3}}
    \institute{
    {Aix Marseille Univ. CNRS, CNES, LAM Marseille, France,}
    \and {National Centre for Nuclear Research, ul. Ho\.za 69, 00-681 Warszawa, Poland, } 
    \and {Astronomy Centre, Department of Physics and Astronomy, University of Sussex, Falmer, Brighton BN1 9QH, UK} 
    \and{Univ Lyon, Univ Lyon1, ENS de Lyon, CNRS, Centre de Recherche Astrophysique de Lyon UMR5574, F-69230, Saint-Genis-Laval, France} 
    \and {Leiden Observatory, Leiden University, NL-2300 RA Leiden, Netherlands} 
    \and {School of Sciences, European University Cyprus, Diogenes Street, Engomi, 1516, Nicosia, Cyprus}
    \and {Department of Physics and Astronomy, University of the Western Cape, Robert Sobukwe Road, Private Bag X17, 7535 Bellville, Cape Town, SA} 
    \and {INAF - Istituto di Radioastronomia, via Gobetti 101, 40129 Bologna, Italy} 
    \and {University of Hawaii, 2505 Correa Road, Honolulu, HI 96822, USA} 
    \and{Department of Astronomy, University of Cape Town, Private Bag X3, Rondebosch 7701, South Africa} }  

  \abstract
   {}
{ 
        The \textit{Herschel} Extragalactic Legacy Project (HELP) focuses on the data from ESA’s \textit{Herschel} mission, which covered over 1300~$\rm deg^2$ and is preparing to publish a multi-wavelength catalogue of millions of objects.
        Our main goal is to find the best approach to simultaneously fitting spectral energy distributions (SEDs) of millions of galaxies across a wide redshift range to obtain  homogeneous estimates of the main physical parameters of detected infrared (IR) galaxies.
}
{       We perform  SED fitting on the ultraviolet(UV)/near-infrared(NIR) to far-infrared(FIR) emission of 42~047 galaxies from the pilot HELP field: ELAIS~N1. 
        To do this we use the latest release of CIGALE, a galaxy SED fitting code relying on energy balance, to deliver the main physical parameters such as stellar mass, star formation rate, and dust luminosity.  
        We implement additional quality criteria to the fits by calculating $\chi^2$ values for the stellar and dust part of the spectra independently. 
        These criteria allow us to identify the best fits and to identify peculiar galaxies. 
        We perform the SED fitting of ELAIS~N1 galaxies by assuming three different dust attenuation laws separately allowing us to test the impact of the assumed law on estimated physical parameters. 
         }
{   We implemented two additional quality value checks for the SED fitting method   based on stellar mass estimation and energy budget.  
        This method allows us to identify possible objects with incorrect matching in the catalogue and peculiar galaxies; we found 351 possible candidates of lensed galaxies using two complementary $\rm \chi^2$s criteria (stellar and infrared  $\rm \chi^2$s) and photometric redshifts calculated for the IR part of the spectrum only. 
        We  find that the attenuation law has an important impact on the stellar mass estimate (on average leading to disparities of a facto33131corrr of two). 
        We derive the relation between stellar mass estimates obtained by three different attenuation laws and we find the best recipe for our sample. 
        We also make independent estimates of the total dust luminosity parameter from stellar emission by fitting the galaxies with and without IR data separately.
}
   {}


\maketitle
   
\section{Introduction}

Multi-wavelength data for extragalactic objects is a necessary precondition for a physical analysis of galaxies, as the full complexity of the galaxy is only seen when using different spectral ranges simultaneously.
The emission from the hot gas component, active nuclear regions, and the end products of stellar evolution (supernovae and compact remnants) can be observed in the X-rays \citep[ 1$<\lambda<$100\AA{}; i.e.][]{2006ARA&A..44..323F}. 
The far-ultraviolet ($\lambda>$912\AA{}; FUV) to infrared ($\sim$3$\mu$m to 1~mm spectral range; IR) spectra of all galaxies arises from stellar light; either directly or reprocessed by the surrounding interstellar medium (ISM). 
Old stars can be seen in the near-infrared (NIR) spectral range (0.75$<\lambda<$5~$\mu$m). 
The dust, composed of a mixture of carbonaceous and amorphous silicate grains \citep{Draine2003ARAA} is heated by the interstellar radiation field and emits in the mid- ($\sim$5$<\lambda<$20~$\mu$m) and far-infrared (MIR and FIR, respectively)\ (10~$\mu$m$<\lambda<$1~mm).

Massive young stars dominate the short wavelength range, while evolved stars mainly emit in the NIR \citep{Kennicutt1998}.   
The dust component, heated by the interstellar radiation field, can be observed in the wavelength range from NIR to submillimetre \cite[i.e. ][]{Hao:2011,daCunha10,Calzetti:2012}. 
For a more detailed review of the multiwavelength emissions of different galaxy components see the book written by  \cite{2011pvg..book.....B}. 

The  UV-to-IR  spectral  energy distribution (SED) contains information about the stars of the galaxy such as the stellar mass or star formation rate (SFR).
For example, information on newborn stars can be inferred from the UV data, making the UV range a very efficient tracer of SFR. 
Unfortunately, one of the obstacles to observing starburst regions in the UV range only is dust.
The newly born stars are surrounded by gas and dust, which, as well as obscuring the most interesting regions,  absorb a part of the UV light emitted by stars \citep[e.g.][]{buat07}.
Dust grains absorb or scatter photons emitted by stars and re-emit the energy over the full IR range. 
Therefwore, only part of the energy from newly born stars can be observed in UV wavelengths.   
Infrared emission, reflecting the dust-obscured star formation activity of galaxies \citep{genzel00}, combined with UV and optical data, can provide a broad range of information about the star formation history (SFH) and SFR.   

As  shown in previous studies \citep[e.g.][]{lefloch05,takeuchi05a}  the fraction of hidden SFR estimated by UV emission increases from $\sim$50\% in the Local Universe to $\sim$80\% at z=1. 
\cite{Burgarella2013} combined the measurements of the UV and IR luminosity functions up to redshift 3.6 to calculate the redshift evolution of the total $\rm SFR_{UV+IR}$ and dust attenuation. 
They found that the dust attenuation increases from z=0 to z=1.2 and then decreases to higher redshifts.  
Therefore the complexity of the SFR seen from UV and IR wavelengths requires the combination of UV and IR data for a proper analysis of the SFH, and its evolution across the history of the universe. 
Moreover, the ratio between the UV and FIR emission serves as an indicator of the dust attenuation in galaxies \citep[e.g.][]{buat05,takeuchi05a}, and the IR wavelengths serve as a tracer for active galactic nuclei (AGNs)  \cite[e.g. ][]{Leja2017}.  
All these factors make the multi-wavelength data set a powerful source of information for detailed galaxy studies.
The most common and effective approach to obtaining constraints on, for example, stellar masses, SFRs, and dust properties from the large multi-wavelength catalogues of galaxies is to fit physical models to the galaxy's broad-band SED  \citep[i.e.][]{Walcher2011,Leja2017}.  
For an overview of recent improvements in models and methods, \citealp{Walcher2011} present a detailed review of different techniques for SED fitting.
Many new tools were developed for both UV-optical data or the dust part only, such as STARLIGHT \citep{STARLIGHT}, ULYSS \citep{ULYSS}, VESPA \citep{VESPA}, Hyperz \citep{hyperz}, Le Phare \citep{arnouts99,Ilbert:2006}, PAHFIT \citep{PAHfit} or to fit specific types of objects \citep[e.g. AGNfitter, ][]{AGNfitter}. 
Also, the Virtual Observatory tools allow for SED fitting. 
Bayesian analysis is included in many different software packages; for example GOSSIP \citep{GOSSIP}, CIGALE \citep{noll09}, and BayeSED \citep{BayeSED}.  
The wealth of public and private SED fitting tools implies that different surveys tend to be analysed with different tools, with no common set of models or parameters.  
As a consequence, it is difficult to combine, compare or interpret large datasets for statistical analysis, as different approaches,  models, and assumptions result in disparate accuracy, scaling factors, and non-uniform physical parameters across a wide redshift range. 

A lack of homogeneous multi-wavelength catalogues  (covering over 10\% of the entire sky), together with non-uniform physical parameters obtained based on different models and software, makes the analysis of the main physical properties of galaxies statistically limited and biased.

An FP7 project called the \textit{Herschel} Extragalactic Legacy Project \citep[HELP,][Oliver et al., in preparation]{Vaccari:2016} funded by  European Union will remove the barriers to multi-$\lambda$ statistical survey science. 
The main aim of HELP is to provide homogeneously calibrated multi-wavelength catalogues covering roughly 1300~$\rm deg^2$ of the \textit{Herschel} Space Observatory \citep{Pilbratt2010} survey fields. 
These catalogues are going to match individual galaxies across broad wavelengths, allowing for multi-wavelength SED fitting to be performed and for statistical studies of the local-to-intermediate redshift galaxy population. 
The selection criteria, depth maps, and master list creation details will be published together with the catalogues (Shirley et al. in prep.).
The presence of IR  data together with UV--NIR counterparts makes the HELP multi-wavelength catalogue a perfect data set for studying galaxy formation and evolution over cosmic time. 

In this paper, we present the general HELP strategy for SED fitting for the millions of galaxies observed in multi-wavelength pass-bands.
We discuss the software, models, and parameters which are going to be used uniformly for each data set. 
This approach ensures homogeneity of obtained physical parameters for the final HELP deliverable of 1300~deg$^{2}$ field. 

The paper is organised as follows. 
In Sect.~\ref{sec:Data} we describe the ELAIS~N1 field as a pilot field of the HELP project.  
In Sect.~\ref{sec:method} we present the method applied in this work to fit SEDs and the main physical models and parameters used for the fitting.  
Section~\ref{sec:Reliability} presents all reliability tests as well as two implemented quality checks. 
In Sect.~\ref{sec:General_properties} we discuss the general properties of the sample, while in  Sects.~\ref{sec:Dust_att} and~\ref{sec:Ldust_prediction}  we explore how all physical parameters obtained by fitting SEDs with different dust attenuation models can be biased and whether or not it is possible to predict total dust luminosity from stellar emission only. 
Our summary and conclusions are then presented in  Sect.~\ref{sec:Conclusions}.
In this paper we use WMAP7 cosmology \citep{Komatsu:2011}: $\rm \Omega_m$=0.272, $\rm \Omega_{\Lambda}$=0.728, $\rm H_0$=70.4~km~s$^{-1}$~Mpc$^{-1}$.

\section{Data}
\label{sec:Data}

\begin{figure*}[th!]
        \begin{center}
                \includegraphics[width=0.92\textwidth]{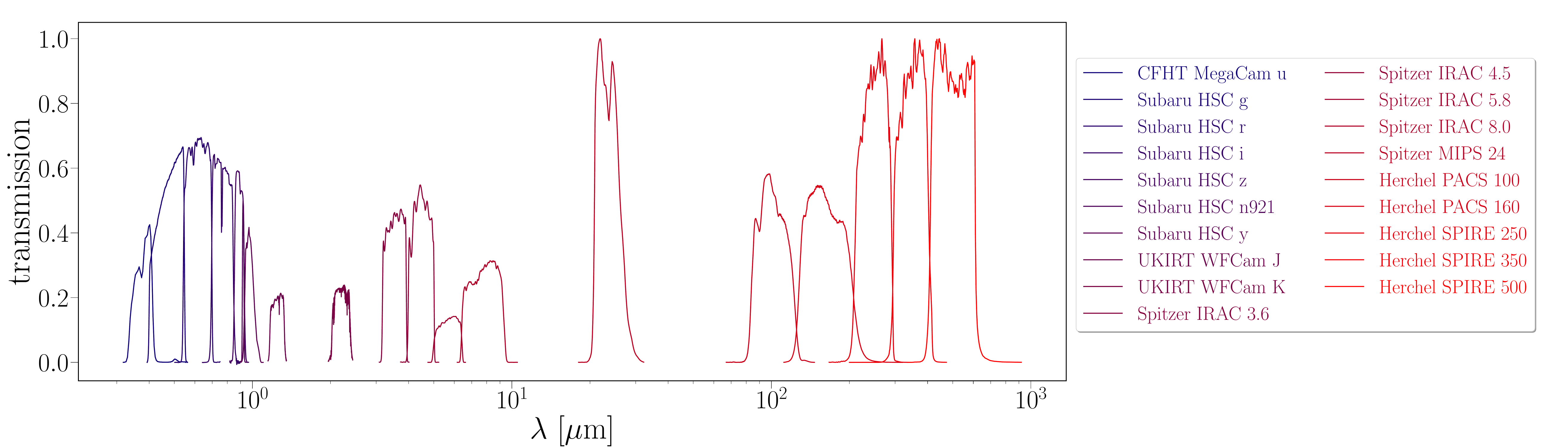}
        \end{center}
        \caption{Transmission curves and demonstrative coverage of primary photometric bands used for SED fitting. 
                Where multiple filters cover the same region, we show only one for clarity. All transmission curves are plotted as exact transmissions. } \label{fig:filters}
\end{figure*}

In our analysis, we focus on the pilot HELP field: European Large Area ISO Survey North~1 (hereafter ELAIS~N1), 9 deg$^2$ area centred at 16$^{h}$10$^{m}$01$^{s}$ +54$^{o}$30$^{'}$36$^{''}$ \citep{Oliver2000}.  
ELAIS~N1 is one of 20 fields making up the European Large Area ISO Survey \citep[ELAIS,][]{Oliver2000,MRR2004}.
The \textit{Herschel} data in ELAIS N1 was obtained as part of the HerMES project \cite{Oliver:2012}. 
It is representative of moderately deep fields for future HELP catalogues. 

In this paper we briefly summarise the data used for ELAIS~N1. 
A detailed description of the data used for the HELP project (both FIR and ancillary data), the open source pipeline, which was developed for HELP, and the cross-matching procedure, astrometry corrections, and full data  diagnostics  will be presented in Shirley et al., (in preparation).

\subsection{Herschel far-infrared sample of Elais~N1 field}
\label{sec:FIRdata}
\textit{Herschel} was equipped with two continuum imaging instruments, the Photodetector Array Camera and Spectrometer \citep[PACS;][]{PACS_Poglitsch2010} and the Spectral and Photometric Imaging Receiver \citep[SPIRE,][]{SPIRE_Griffin2010}. 
These instruments provided FIR coverage at 100 and 160~$\mu$m from PACS, and at 250, 350, and 500~$\mu$m from SPIRE.
To obtain the photometry of \textit{Herschel} sources a new prior-based source extraction tool was developed, called XID+ \citep{Hurley2017}. 

The XID+\footnote{The software is available at  \href{https://github.com/H-E-L-P/XID_plus}{https://github.com/H-E-L-P/XID\_plus}}  
is a probabilistic de-blending tool used to extract \textit{Herschel} SPIRE source flux densities from \textit{Herschel}  maps that suffer from source confusion \citep{Nguyen:2010}. 
This is achieved by using a Bayesian inference tool to explore the posterior probability distribution. 
This algorithm is efficient in obtaining Bayesian 
probability distribution functions (PDFs) for all prior  sources,  and  thus  flux  uncertainties  can  be  estimated. 
A detailed description can be found in \cite{Hurley2017}. 
The original XID+, used for ELAIS~N1 field, uses information IRAC bands as a prior. 
We  note that XID+  can  be  run  with  more  sophisticated  priors,  using both dust luminosity and  redshift  information \citep[e.g.][developed  a method of incorporating flux predictions from SED fitting procedure as informed priors, finding improvements in the detection of faint sources]{Pearson:2018}. 
The  XID+ was run on the ELAIS~N1 Spitzer MIPS, and \textit{Herschel} PACS and SPIRE map. 
The flux level, at which the average posterior probability distribution of the source flux becomes Gaussian \citep[indicating that the information from data dominates over the prior,  see] for more details]{Hurley2017}, is 20~mJy for MIPS, 12.5 and 17.5~mJy for PACS and for SPIRE is 4~$\mu$Jy for all three bands.

In the deblending work, the priors we use for computing XID+ fluxes must satisfy two criteria: they must have a detection in the Spitzer IRAC 1 band and they must have been detected in either optical of NIR (this was done to eliminate artefacts). 
The entire catalogue of the HELP FIR measurements based on the XID+ tool will be published at the end of the program (2018), together with the full multi-wavelength data collected from other surveys.

\subsection{Ancillary data}

The HELP master catalogue is built on a positional cross match of all the public survey data available in the optical to MIR range. 
This comprises observations from the Isaac Newton Telescope/Wide Field Camera (INT/WFC) survey \citep{Gonzales-Solares:2011}, the Subaru Telescope/Hyper Suprime-Cam Strategic Program Catalogues (HSC-SSP) \citep{Aihara:2018}, the Spitzer Adaptation of the Red-sequence Cluster Survey (SpARCS) \citep{Tudorica:2017}, the UK Infrared Telescope Deep Sky Survey - Deep Extragalactic Survey (UKIDSS-DXS) 
\citep{Swinbank:2017,Lawrence:2007}, the Spitzer Extragalactic Representative Volume Survey \cite[SERVS, ][]{Mauduit:2012}, and the Spitzer Wide InfraRed Extragalatic survey  \citep[SWIRE, ][]{Lonsdale:2003, Surace:2005}. 
We use the Spitzer Data Fusion products\footnote{\url{http://www.mattiavaccari.net/df/}} for the final two Spitzer `MIR surveys presented in \cite{Vaccari:2010} and \cite{Vaccari2015}. 
The cross match is described in full in Shirley et al. (in prep). 
The list of filters is shown in Table~\ref{tab:ancillary_data_tab} and the coverage is presented in Fig.~\ref{fig:filters}. 

When multiple measurements are available in similar passbands we take the deepest only since rapidly increasing errors on shallower surveys mean those measurements contribute little to overall photometry. 
In cases where a detection is available from two similar filters (e.g. MegaCam~g and HyperSuprimeCam~g) we check the number of measurements in the full catalogue, the depth associated with each filter, and the distribution of uncertainties.  
Because there  is a large difference in the depths of different surveys, there is no advantage to using multiple measurements where one clearly has an order-of-magnitude lower error and errors in the shallower catalogues may not include systematic errors of comparable size to the random errors.
Based on our analysis we decide to use filters in the same order for each band: HyperSuprimeCam -- MegaCam -- WFC -- GPC1.
This means that if we have, for example, g band measurements from WFC and MegaCam, for our analysis, we are going to use only measurements from MegaCam.  If we also have   HyperSuprimeCam photometry then only that one would be used for SED fitting.

We also remove objects around bright stars by measuring the size of the circular region around a star that contains no detections as a function of the star magnitude. 
This compound selection function can be used to model which objects will propagate through to the final SED sample.

{\renewcommand{\arraystretch}{1.3}
\begin{table}[]
        \begin{center} 
                \caption[]{Data used for SED fitting from ELAIS~N1 field. }
                \label{tab:ancillary_data_tab}
                \begin{tabular}{l| l| l} \hline  
                        Telescope & Instrument &  Filter\\ \hline
                        CFHT & MegaCam &  u, g, r, y, z\\
                        Subaru &  HSC & g, r, i, z, N921, y\\
                        Isaac Newton  &  Wide Field Cam. &  u, g, r, i, z\\
                        PanSTARRS1 & Gigapixel Cam.1 & g, r, i, z, y\\
                        UKIRT & WFCam & J, K\\
                        {\multirow{2}{*}{Spitzer} } & IRAC & 3.6, 4.5, 5.8, 8.0 ($\mu$m) \\
                        & MIPS & mips~24~($\mu$m)\\
                        {\multirow{2}{*}{\textit{Herschel}} } & PACS & 100, 160 ($\mu$m)\\
                        & SPIRE & 250, 350, 500 ($\mu$m)\\ \hline               
                \end{tabular}
        \end{center}
\end{table}
        
\begin{figure}[ht!]
        \begin{center}
                \includegraphics[width=0.41\textwidth]{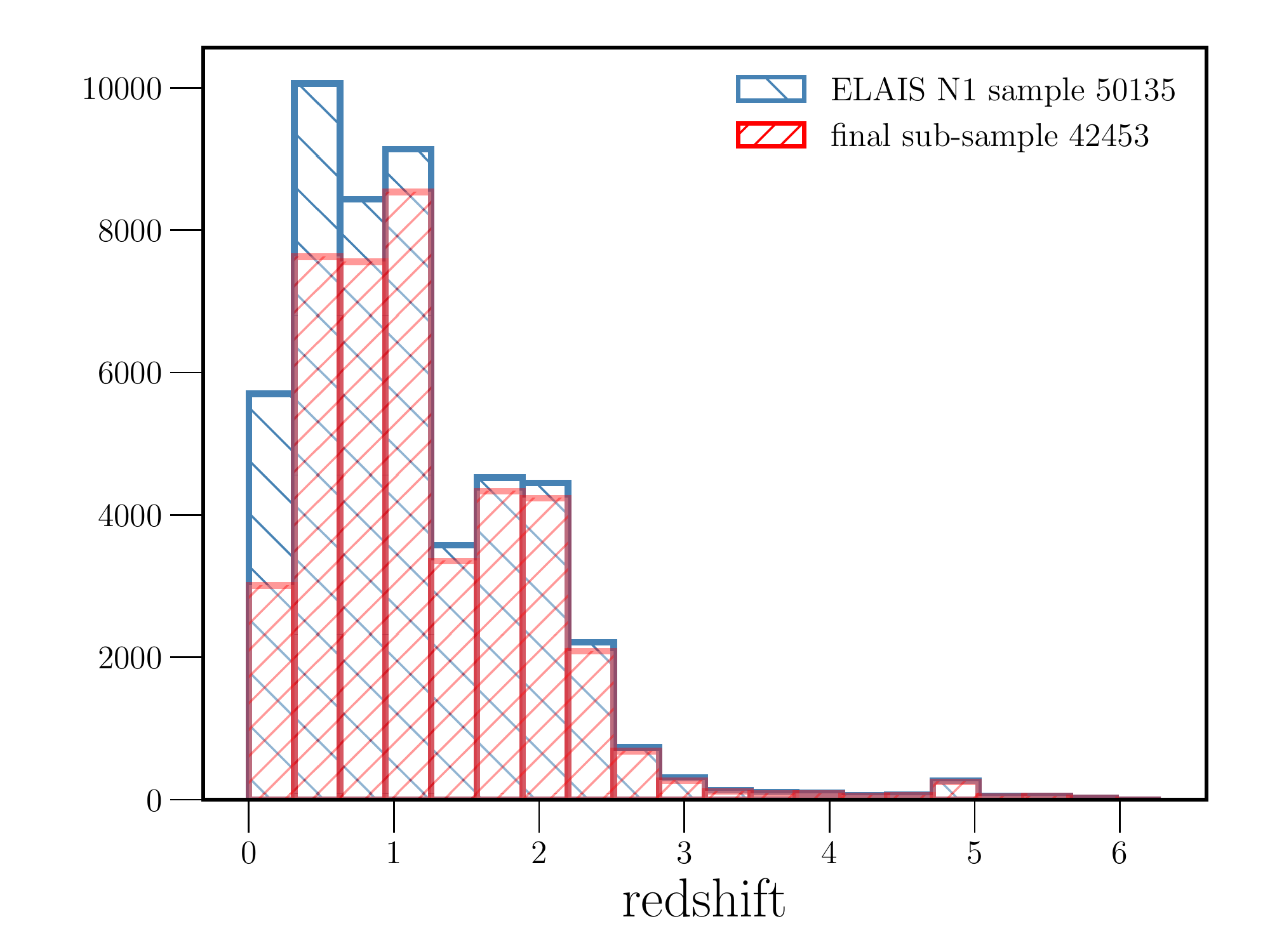}
        \end{center}
        \caption{Redshift distributions of 50~129 galaxies for which is was possible to fit SEDs (blue histogram) and { 42~453} galaxies remaining after SED quality cleaning according to Sect.~\ref{sec:twochi2s} (red histogram). } \label{fig:redshift}
\end{figure}
        
\subsection{Photometric redshifts}
        
As part of the HELP database, we provide new photometric redshifts generated using a Bayesian combination approach, described in  \cite{Duncan2017}. 
They used two independent multiwavelength datasets (NOAO Deep Wide Field Survey Bootes and COSMOS) and performed \textit{zphot} estimation. 
\cite{Duncan2017} investigated the performance of three \textit{zphot} template  sets: (1) default EZY reduced galaxy set \citep{Brammer2008}, (2) "XMM COSMOS" templates \citep{Salvato2009}, and (3) atlas of Galaxy SEDs \citep{Brown2014}  as  a  function  of  redshift,  radio  luminosity,  and  IR/X-ray  properties.
They found that only a combination of all template libraries is able to provide a consensus \textit{zphot} estimate and they used a hierarchical model Bayesian combination of the \textit{zphot} estimates.   
        
This method is used for the HELP project and for the next generation of deep radio continuum surveys. 
The detailed description of the \textit{zphot} methodology and redshift accuracy is presented in \cite{Duncan2017}.

\subsection{Final sample}
\label{sec:final_sample}
        
The catalogues produced by XID+ contain a flag for those sources that are either below a flux level in which there is a clear detection or the related XID+ Bayesian P-value maps indicate a poor fit in a region local to the source.
The final sample includes  50~129 galaxies from the ELAIS~N1 survey with flux measurements for  PACS or SPIRE data (or both). 
The full sample and data files can be downloaded at \url{http://hedam.lam.fr/HELP/dataproducts/dmu28/dmu28_ELAIS-N1/data/zphot/} and also accessed with Virtual Observatory standard protocols at \url{https://herschel-vos.phys.sussex.ac.uk/}.
        
Our procedure gives us a set of 19 bands: u, g, r, i, z, N921, y, J, K, Spitzer IRAC 3.6, 4.5, 5.8 and 8.0~$\mu$m, Spitzer MIPS 24~$\mu$m, and  five passbands from \textit{Herschel} PACS (100 and 160~$\mu$m) and SPIRE (250, 350 and 500~$\mu$m).
The coverage of the wavelengths is shown in Fig.~\ref{fig:filters}.
        
The mean value of the redshift distribution of the final sample is 0.97, while 50\% of galaxies (between quartile 1 and 3) are located in the redshift range 0.53 -- 1.63.  
The redshift distribution of the initial sample can be found in Fig.~\ref{fig:redshift} (blue hatched histogram).

\section{Fit of the spectral energy distribution}
\label{sec:method}

Taking advantage of the very dense coverage from the broadband passbands (from u band to 500~$\mu$m), we use Code Investigating GALaxy Emission\footnote{\url{http:\\cigale.lam.fr}} (CIGALE, Boquien et al., in preparation).
CIGALE is designed to estimate the physical parameters (i.e. SFR, stellar mass, dust luminosity, dust attenuation, AGN fraction)  by comparing modelled galaxy SEDs to observed ones. 
CIGALE conserves the energy balance  between the dust-absorbed stellar emission and its re-emission in the IR. 
Many authors \citep[e.g.][]{Buat15,Ciesla15,Ciesla2016,Malek2017} have presented the methodology and the strategy of the code already. 
A more detailed description of the code will be given by Boquien et al (in prep).
All adopted parameters used for all modules are presented in Table.~\ref{tab:input}.

\subsection{Stellar component}

To build the stellar component, we use  the stellar population synthesis models by \cite{bruzal03} with the initial mass function (IMF) given by \cite{Chabrier2003IMF}. 

As the aim of this analysis is to show how to fit the SEDs of a large sample of galaxies, simplicity and limiting the number of parameters is crucial. 
It was shown in  \cite{Ciesla15}, than a SFH composed of a delayed form to model the bulk of the stellar population with the addition of a flexibility in the recent SFH  provides very good estimates of the SFR--${\rm M_{star}}$ relation comparing to observations. 
We refer the reader to \cite{Ciesla15} for a detailed description of the model.  
We apply SFH scenarios which include delayed SFR with an additional burst. 
Parameters which describe our scenario are: age of the galaxy, decreasing rate, burst fraction, and burst age. 
The functional form for the SFR is calculated as:
\begin{equation}
\rm SFR(t)= SFR_{delayd}(t)+SFR_{burst}(t),
\end{equation}
where $\rm SFR_{delayd}(t) \propto t e^{{-t}/{\tau_{main}}}$, and $\rm SFR_{burst}(t) \propto t e^{{-(t-t_0)}/{\tau_{burst}}}$ if $t>t_0$, and $\rm SFR_{burst}(t)=0$ if $t<t_0$. The factor 
$\tau_{main}$ represents e-folding time of the main stellar population and $\tau_{burst}$ represents e-folding time of the late starburst population.
The adopted parameters used for the stellar component in our fitting procedure are presented in Table~\ref{tab:input}. 
We fix values of e-folding times of the main and late stellar population models as those parameters are very difficult to constrain using the SED fitting procedure \citep[i.e.][]{noll09}. }

\subsection{Infrared emission from dust}

The current public version of CIGALE (cigale version 0.12.1) includes four different modules to  calculate  dust  properties: \cite{casey12},  \cite{dale07}, \cite{dale14}, and an updated version of the \cite{draine07} model, which we decide to use  (this model was also used by e.g. \citealp{LoFaro2017} and \citealp{Pearson:2018} in the framework of the HELP project). 
In our analysis we apply the multi-parameter dust emission model proposed by \cite{Draine:2014}, which is described as a mixture of carbonaceous and amorphous silicate grains. 
Infrared emission SEDs have been calculated for dust grains heated by starlight for various distributions of starlight intensities. 
The majority of the dust is heated by a radiation field with constant intensity (marked as $\rm U_{min}$ in Table~\ref{tab:input}) from the diffuse ISM. 
A much smaller fraction of dust ($\gamma$, from Table~\ref{tab:input}) is illuminated by the starlight with intensity range from $\rm U_{min}$ to $\rm U_{max}$. 
This intensity is characterised by a power-law distribution.

We note that we are not able to obtain a reliable value of the $\gamma$ parameter due to the degeneracy between $\gamma$ and radiation intensity, and the large photometric uncertainty.
We fix the value of $\gamma$ using the mean value obtained from a stacking analysis by \cite{Magdis2012} for the average SEDs of main sequence galaxies at redshifts 1 and 2 ($\gamma$=0.02).

\subsection{AGN component}

As shown in the literature \citep[e.g.][]{Ciesla15,Leja2018} the SED fitting procedure is  a powerful technique, but the accuracy of estimated physical properties  is tightly correlated with  the accuracy of the models used.
For example, the AGNs  can substantially contribute to the MIR emission of a galaxy. 

To improve the derived galaxy properties we add an AGN component to the stellar SED. 
We derive the fractional contribution of the AGN  emission from \cite{fritz06} templates, which assume  two components: (1) point-like isotropic emission of the central source, and (2) radiation from dust with a toroidal geometry in the vicinity of the central engine. 
The AGN emission is absorbed by the toroidal obscurer and re-emitted at 1--1000~$\rm{\mu}m$ wavelengths or scattered by the same obscurer. 
We perform SED fitting with a set of parameters from the \cite{fritz06} models as described in Table~\ref{tabela_par}. 
To limit the number of models we fix the value to five variables that parametrizes the density distribution of the dust within the torus (ratio of the maximum to minimum radius of the dust torus, radial and angular dust distribution in the torus, angular opening angle of the torus, and angle between equatorial axis and the line of sight) with typical values found by \cite{fritz06}, and used, for example, by \cite{Hatziminaoglou09}, \cite{Buat15}, and \cite{Ciesla15}.

\subsection{Dust attenuation} 

We perform three SED-fitting runs with three different dust-attenuation laws to check  which law gives the best fits: we use the \cite{CF00} model as the main dust attenuation recipe, and then we redo the whole analysis with the popular attenuation law of \cite{calzetti00}. Subsequently we also test the attenuation law for z$\sim$2 ULIRGs  derived by \cite{LoFaro2017}. 

The standard formalism of \citealt{CF00} (hereafter:~\citetalias{CF00}) assumed that stars are formed in interstellar birth clouds (BCs), and after $\rm 10^7$~yr, young stars disrupt their BCs and migrate into the ambient ISM.
The functional form for \cite{CF00} attenuation law  (CIGALE's \texttt{dustatt\_2powerlaws} module) refers to: 
\begin{equation}
\label{eq:CF00}
\rm A({\lambda}) = \begin{cases} 
\rm A_{\lambda}(BC) + A_{\lambda}(ISM)  \text{,  for young stars, age $<10^{7}$yr},\\ \rm A_{\lambda}(ISM)   \text{,  for old stars (age$>10^7$yr) }. \\
\end{cases} 
\end{equation}
where $\rm A_{\lambda}=A_V(\lambda/\lambda_V)^{\delta}$, $\rm \lambda_V$ is 550 $\mu$m, $\rm A_V$ is a V-band attenuation, and  for \citetalias{CF00} $\delta$=-0.7.
The fraction of the total effective optical depth contributed by the diffuse ISM is defined as  $\rm \mu=f_{att}/(1+f_{att})$, where $\rm f_{att}=A^{ISM}_{V}/A^{BC}_{V}$. 

As part of the HELP project, \citealp{LoFaro2017} compared power-law attenuation curves with the results of radiative--transfer calculations for 20 ULIRGs observed by \textit{Herschel}. 
\cite{LoFaro2017} provided an attenuation law similar to but flatter than that of \citetalias{CF00},  which consists of two power laws but with slopes of the attenuation in the BC and ISM equal to -0.48. 
The effect of the change of the slopes from -0.7 to -0.48 results in ever more greyer attenuation curves for the UV part of the spectrum. 
The functional form of \cite{LoFaro2017} attenuation law is exactly the same as Eq.~\ref{eq:CF00} with  $\delta$=-0.48.

The functional form for \cite{calzetti00} attenuation law is described as: 
\begin{equation}
\label{eq:Calzetti}
\rm A(\lambda) =  E(B-V)k(\lambda)
,\end{equation}
where $\rm k(\lambda)$ is the effective attenuation curve  defined as:
\begin{equation}
\rm k(\lambda)=\begin{cases}
\rm 2.659(-1.857+1.040/\lambda), \\ 
\text{ for }(0.63\mu m \leqslant \lambda \leqslant2.20 \mu m)\\
\rm 2.659(-2.156+1.509/\lambda-0.198/\lambda^2+0.011/\lambda^3), \\ \text{ for } (0.12\mu m \leqslant \lambda < 0.63 \mu m),
\end{cases} 
\end{equation}  
and $\rm E(B-V)$ is the colour excess for the stellar continuum. 

The difference in shape for attenuation curves defined by  \cite{calzetti00} and \citetalias{CF00} is mostly present for the wavelengths longer than 5000~$\AA$, when \citetalias{CF00} curve is much flatter than the one given by \cite{calzetti00}. 
A detailed description of the difference between the two attenuation laws can be found in \cite{LoFaro2017}. 

To make the comparison possible, we used only one 
reduction factor between $\rm A^{ISM}_{V}/A^{BC}_{V}$  ($\rm f_{att}$)  both for \citetalias{CF00} and \cite{LoFaro2017}. 
As was shown by \cite{LoFaro2017} and references therein, the $\rm f_{att}$ parameter is known to usually be unconstrained by broad-band SED fitting.   
We performed a preliminary analysis using $\rm f_{att}$ as a free parameter (values correspond to $\mu$ parameter in the interval 0.2–0.5), and for a final run we chose the most probable value obtained based on the full ELAIS~N1 sample: 0.44 ($\rm f_{att}=0.8$). 

To summarise, the basic attenuation law we used is that of \citetalias{CF00}. 
Then we compared our results with  the \cite{calzetti00} recipe to check our results with those widely presented in the literature, and then with the \cite{LoFaro2017} law for ULIRGs.
In the next step we  performed the comparison between results obtained with all three attenuation laws (see Sect.~\ref{sec:Dust_att}).

{\renewcommand{\arraystretch}{1.09}
\begin{table*}[]
        \begin{center} 
                \caption[]{List of the input parameters of the code CIGALE. All free parameters are marked with a star  before the name. }
                \label{tabela_par}
                \begin{tabular}{p{0.56\linewidth}| p{0.44\linewidth}} \hline \hline  
                        \label{tab:input}
                        \multirow{2}{*}{parameter} & \multirow{2}{*}{values} \\ 
                        &  \\ \hline 
                        \multicolumn{2}{c}{delayed star formation history + additional burst} \\    \hline \hline
                        e-folding time of the main stellar population model [Myr] & 3000\\
                        e-folding time of the late starburst population model [Myr] & 10 000\\
                        $\star$ mass fraction of the late burst population & 0.001, 0.010, 0.030, 0.100, 0.300\\
                        \multirow{2}{*}{$\star$  age [Myr]}    &  1000, 2000, 3000, 4000, 5000, 6500, 10000\\
                        $\star$  age of the late burst [Myr] & 10.0, 40.0, 70.0\\
                        \hline
                        \multicolumn{2}{c}{single stellar population:  \cite{bruzal03}} \\    \hline \hline
                        initial mass function & \cite{Chabrier2003IMF} \\
                        metallicities (solar metallicity) & 0.20 \\
                        age of the separation between the young and the old star population [Myr] & 10 \\ \hline 
                        \multicolumn{2}{c}{attenuation curve} \\ \hline      \hline 
                        \multicolumn{2}{c}{main recipe: \cite{CF00} } \\ \hline 
                        $\star$  $\rm A_V$ in the BCs & 0.3, 0.8,1.2,1.7,2.3,2.8,3.3, 3.8\\   
                        power law slopes of the attenuation in the birth clouds and ISM &  -0.7 \\
                        \hline
                        \multicolumn{2}{c}{\cite{LoFaro2017} } \\ \hline         
                        $\star$ V-band attenuation in the birth clouds (Av BC) & 0.3, 0.8,1.2,1.7,2.3,2.8,3.3, 3.8\\   
                        power law slopes  of the attenuation in the birth clouds and ISM &  -0.48 \\ 
                        \hline
                        \multicolumn{2}{c}{\cite{calzetti00} } \\ \hline         
                        $\star$  the colour excess of the stellar continuum light for the young population &  0.6, 0.67, 0.74, 0.81, 0.88, 0.95, 1.02, 1.09, 1.16, 1.23, 1.29, 1.36, 1.43, 1.50, 1.57, 1.64, 1.71, 1.78, 1.85, 1.92, 1.99, 2.06, 2.13, 2.2\\ \hline
                        \multicolumn{2}{c}{dust emission: \cite{draine07}} \\                      \multicolumn{2}{c}{for our analysis we built templates  based on adopted parameters from previous studies:}\\ 
                        \multicolumn{2}{c}{\cite{Magdis2012,Ciesla15, LoFaro2017,Pearson:2018}} \\    \hline \hline
                        $\star$  mass fraction of PAH &  1.12, 2.5, 3.19\\
                        $\star$  minimum radiation field ($\rm U_{min}$) & 5.0, 10.0, 25.0\\
                        $\star$ power law slope dU/dM  ($U^{\alpha}$) &  2.0, 2.8 \\
                        fraction illuminated from Umin to Umax ($\gamma$)&  0.02  \\    \hline
                        \multicolumn{2}{c}{AGN emission: \cite{fritz06} }\\
                        \multicolumn{2}{c}{For our analysis we used templates built based on average parameters from previous studies:}\\ 
                        \multicolumn{2}{c}{\cite{fritz06,Hatziminaoglou09,Buat15,Ciesla15,Malek2017}} \\    \hline \hline
                        ratio of the maximum to minimum radius of the dust torus & 60 \\
                        $\star$ optical depth at 9.7 microns & 1.0, 6.0\\
                        radial dust distribution in the torus & -0.5 \\
                        angular dust distribution in the torus & 0.0 \\
                        angular opening angle of the torus [deg]& 100.0 \\
                        angle between equatorial axis and the line of sight [deg] & 0.001\\
                        $\star$ fractional contribution of AGN & 0.0, 0.15, 0.25, 0.8\\ \hline
                \end{tabular} 
        \end{center}
\end{table*}
}

\subsection{Summary of SED templates}

Each created SED template consists of five components (SFH, single stellar population, dust attenuation and emission, and the AGN component) which are spread of all possible grid of the input parameters.

All these parameters (with the exception of parameters related to the attenuation curve recipes published by  \citealp{calzetti00} and  \citealp{LoFaro2017}, which were used for comparing the main physical parameters presented in Sect.~\ref{sec:Dust_att}) are default parameters used for all HELP fields.
The full list of the input parameters of the code is presented in Table~\ref{tabela_par}.

Based on the number of parameters presented in Table~\ref{tab:input} and the redshift range for our sample of galaxies  we built $\sim$850 million SED templates which were fitted  with CIGALE to 50~129 ELAIS~N1 galaxies detected by \textit{Herschel}. 
We used a 20-core personal computer (PC) with 252Gb of memory. 
The total time to fit all SEDs for ELAIS~N1 was $\sim$6~hours.  

\section{Reliability check for SED fitting procedure and identification of outliers}
\label{sec:Reliability}
\begin{figure*}[h!]
        \centering
        \includegraphics[width=0.95\textwidth, clip]{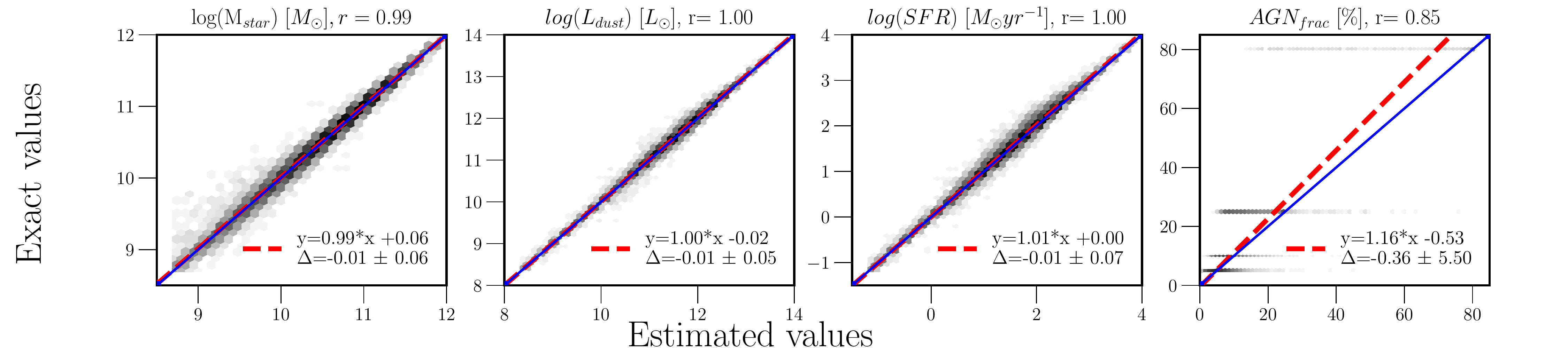}          
        \includegraphics[width=0.95\textwidth, clip]{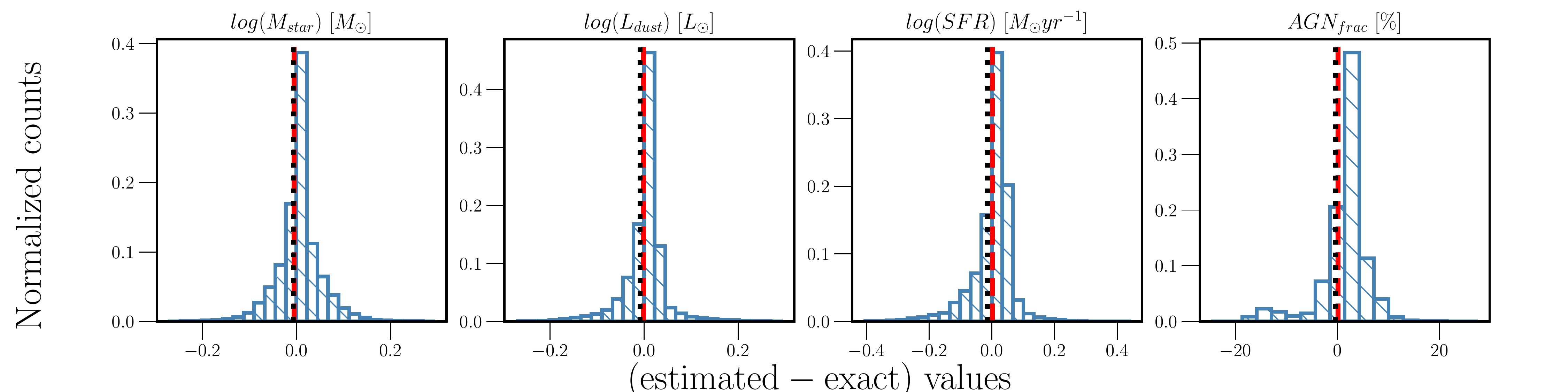}  
        \caption{
        \textit{Upper panel: }Comparison between the true value of the output parameter provided by the  best-fit model for the mock catalogue (x-axis) and the value estimated by the code (y-axis),  for $\rm M_{star}$, $\rm L_{dust}$,  SFR and AGN fraction. 
        The Pearson product-moment correlation coefficient is given as an `r' value. The blue line corresponds to the 1:1 relation, while the red dashed line is a regression line with the equation given in the legend. 
        $\Delta$ represents the mean difference between estimated and exact values and the standard deviation of that difference. 
        \textit{Bottom panel:} distribution of estimated minus exact parameters from upper panel.  Red dashed lines correspond to median values, while black dotted lines represent mean values.      }
                        
        \label{fig:mock}
\end{figure*}

After the SED fitting we removed
143 SED fitting failures (galaxies with $\chi^2_r$ equal to 99\footnote{
The most common reason for SED fitting failures is overestimated redshifts
which translates into an  estimated age of the galaxy larger than the age
of the universe with the assumed ages provided in the CIGALE SED fitting.}) from the initial sample of 50~129 ELAIS~N1. 
From now on we use the remaining 49~986 galaxies as an ELAIS~N1 sample.  

\subsection{Mock analysis}
\label{sec:mock_analysis}

A mock catalogue was generated to check the reliability of the computed physical parameters. 
To perform this test, we use an option included in CIGALE, which allows for the creation of a mock object for each galaxy for which the physical parameters are known. 
To build the artificial catalogue we use the best-fit model for each galaxy used for SED fitting (one artificial object per galaxy). 
A detailed description of the mock analysis can be found in \cite{LoFaro2017} and \cite{giovannoli11}.
In the next step we perturb the fluxes obtained from the best SEDs according to a Gaussian distribution with $\sigma$ corresponding to the observed uncertainty for each band.  
In the final step of the verification of estimated parameters we run CIGALE on the simulated sample using the same set of input parameters as for the original catalogue and compare the output physical parameters of the artificial catalogue with the real ones. 
A similar reliability check was performed for example by \citet{daCunha08}, \cite{Walcher2011}, \cite{yuan11}, \cite{boquien12}, \cite{buat14}, \cite{BayeSED} and  \cite{Ciesla15}.

All the physical parameters presented in Fig.~\ref{fig:mock}  are computed from their probability distribution function (PDF)  as the mean and standard deviation determined from the PDFs  (Boquien~et~al., in prep., or \cite{Walcher08} for a more detailed explanation of the  PDF method.

The upper panel of Fig.~\ref{fig:mock} presents the comparison of the output parameters of the mock catalogue  with the  best values estimated by the code for our real galaxy sample. 
The value, which characterises the reliability of the obtained properties, is the Pearson product-moment correlation coefficient (\texttt{r}). 
The dispersion of the  main physical parameters is presented in the same plot as the $\Delta$ value.
The lower panel of Fig.~\ref{fig:mock} shows the distribution of estimated  minus exact values for $\rm M_{star}$, $\rm L_{dust}$, SFR, and $\rm AGN_{frac}$.   
We find normal distributions with small ($\sim$0.1) overestimations of $\rm log(M_{star})$ and  $\rm log(L_{dust})$.
The third histogram suggests a greater overestimation of SFR which is the result of the slight overestimation of $\rm L_{dust}$ which propagates to SFR.  
The estimation of the $\rm AGN_{frac}$ is less accurate ($\Delta=-0.36\pm5.50)$ due to  the grid parameters used for a fraction of the AGNs (we were not able to put more than four parameters due to the number of models that needed to be created and the time required to analyse millions of galaxies with all models).

\begin{figure}[ht!]
        \begin{center}
                \subfloat[]
                {\includegraphics[width=0.45\textwidth]{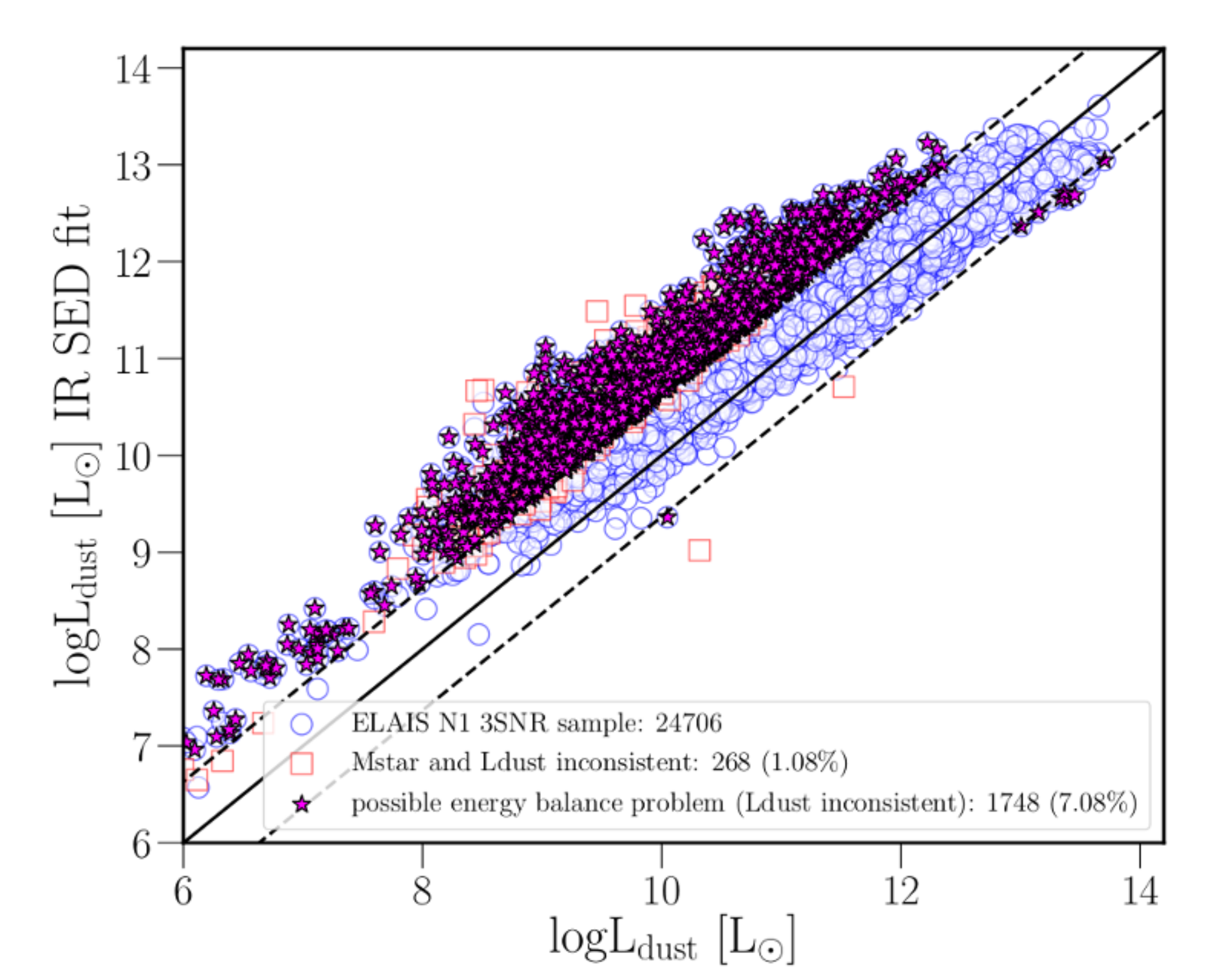}} 
                        
                \subfloat[]
                {\includegraphics[width=0.45\textwidth]{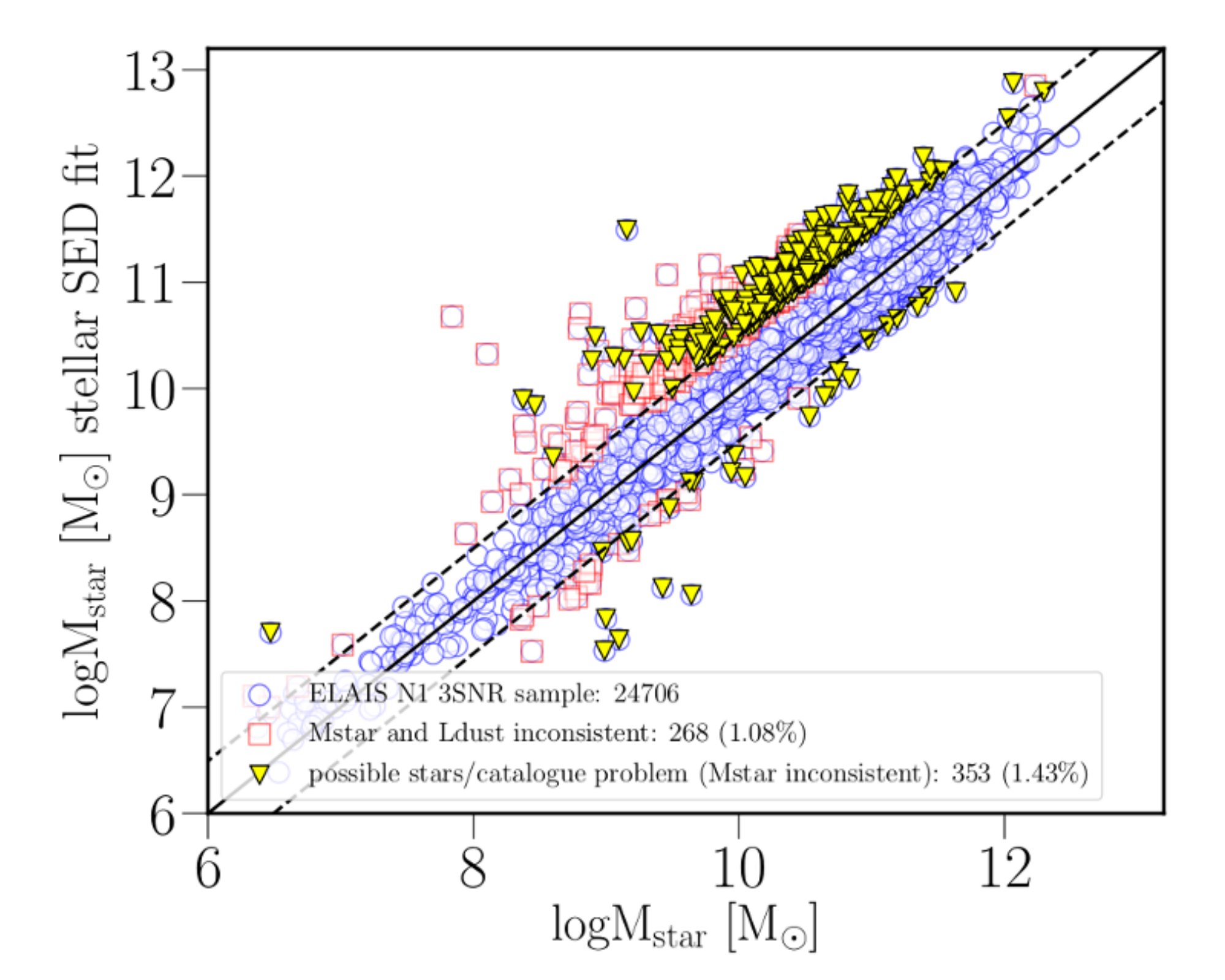}}
        \end{center}
        \caption{Comparison between $\rm L_{dust, UV-FIR\mbox{ }SED} $  and $\rm L_{dust, IR\mbox{ }SED} $ \textit{(panel a)} and $\rm M_{star,UV-FIR\mbox{ }SED}$  and  $\rm M_{star,stellar\mbox{ }SED}$  \textit{(panel b)}. The solid black line represents 1:1 relation, while black dashed lines correspond to 3$\sigma$. Open blue circles  show the ELAIS N1 sample, magenta stars (in panel a) show possible energy budget outliers, filled yellow triangles (panel b) show stellar mass outliers,  and open red squares (panels a and b) show objects with inconsistencies in both physical parameters ($\rm L_{dust}$ and $\rm M_{star}$).} 
        \label{fig:3sigma}
\end{figure}

\begin{figure*}[ht!]
        \begin{center}
                \subfloat[]
                {\includegraphics[width=0.33\textwidth]{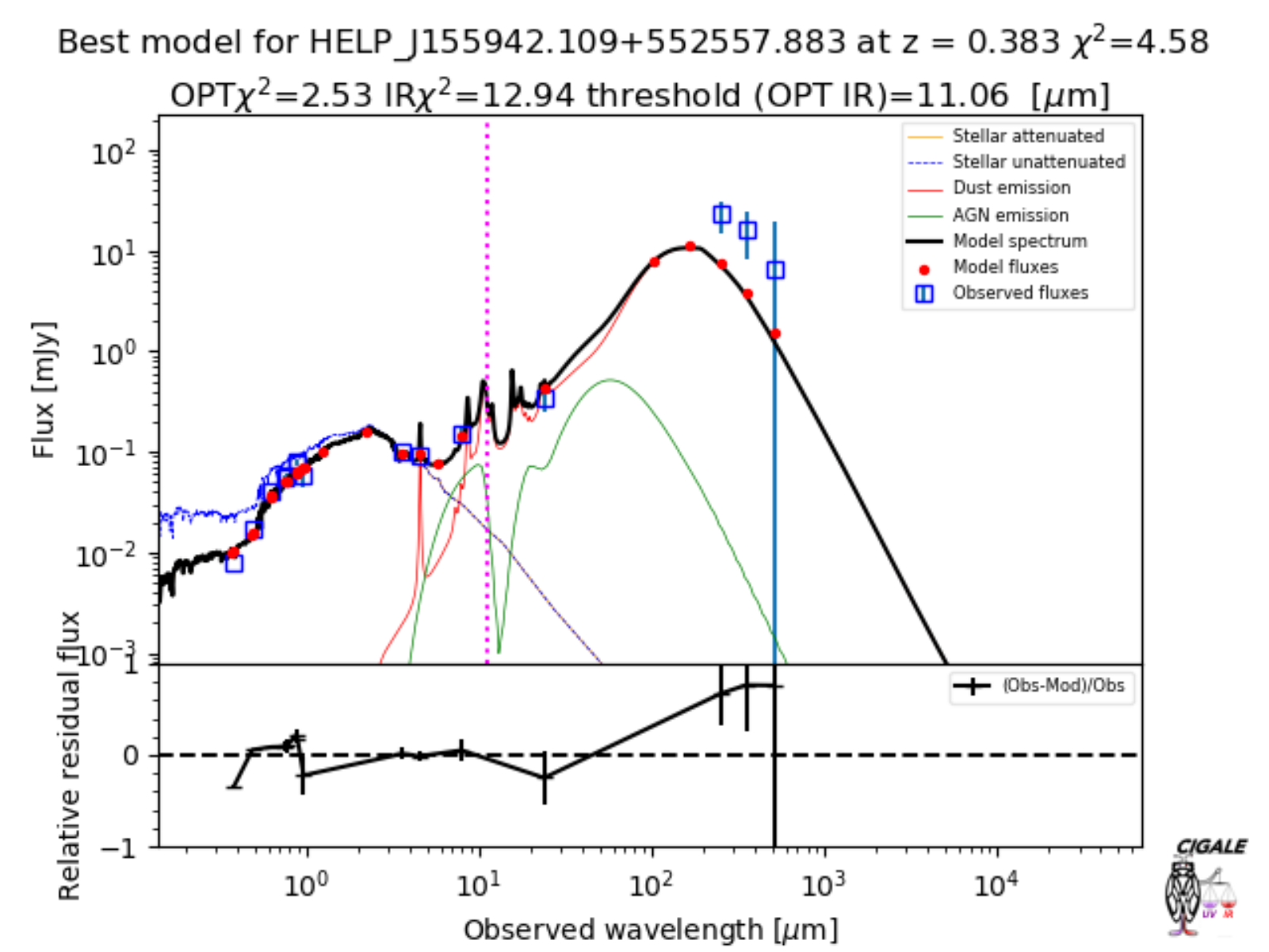}} 
                \hspace{\fill}
                \subfloat[]
                {\includegraphics[width=0.33\textwidth]{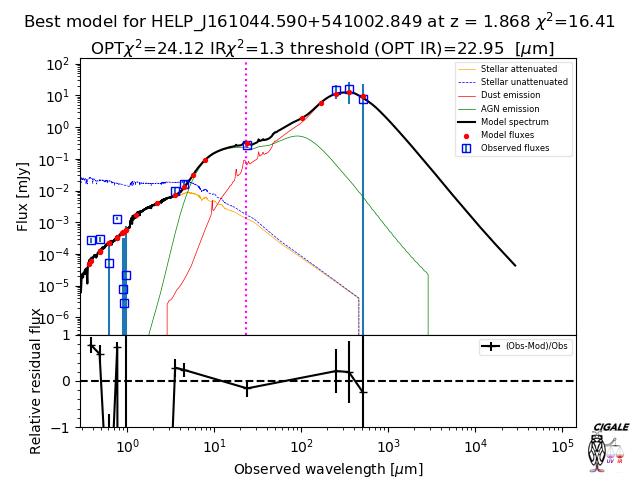}}
                \hspace{\fill}
                \subfloat[] {\includegraphics[width=0.33\textwidth]{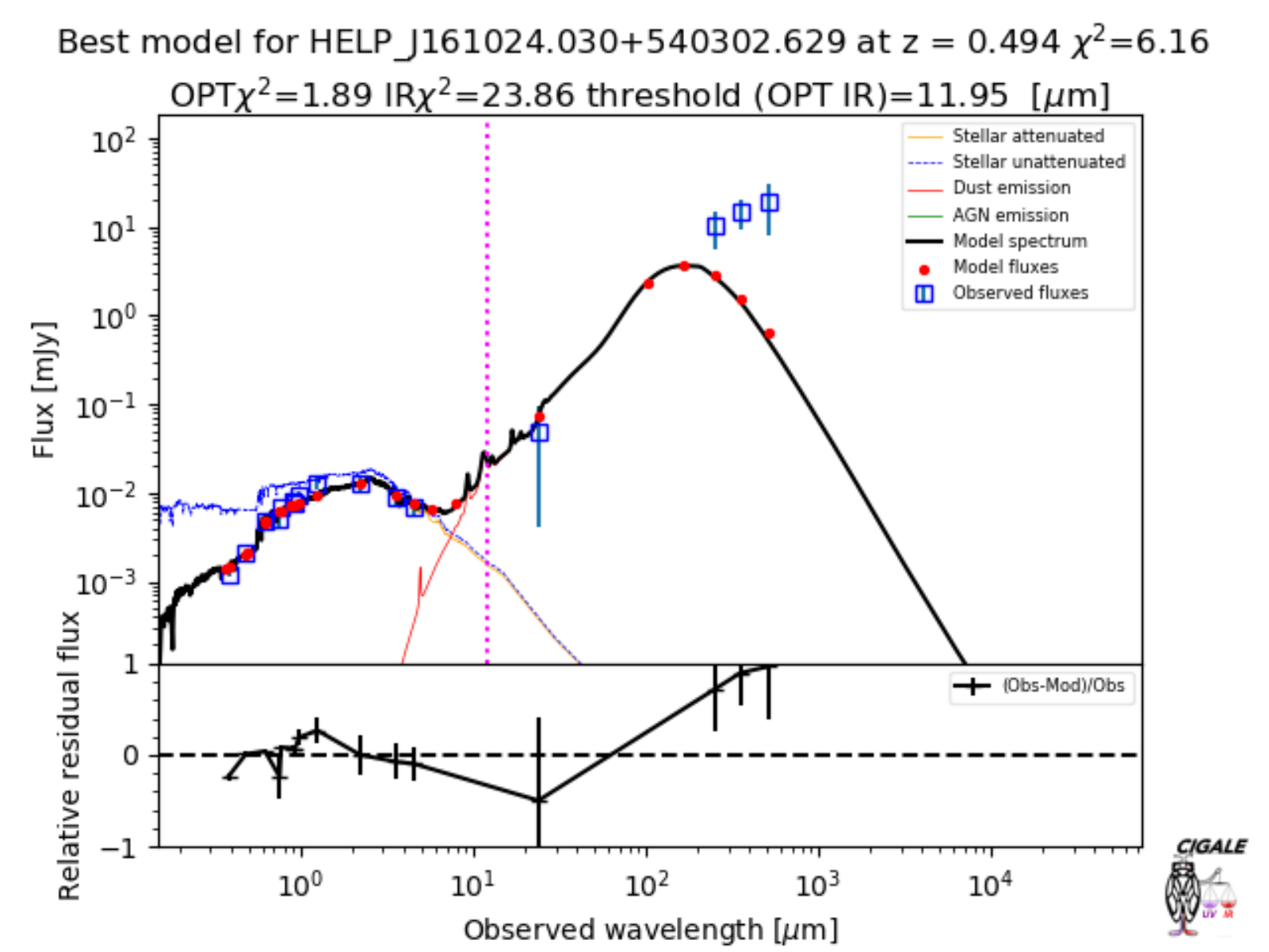}}
        \end{center}
        \caption{An example of (a) possible energy budget issue, (b) incorrectly matched optical data, (c)  lensed candidate found using a disagreement between stellar and infrared $\rm \chi^2_r$ of the SED fitting. The optical part of the spectrum is fitted very well when $\rm \chi^{2}_{r,IR}$ value is on the tail of the $\rm \chi^{2}_{r,IR}$ distribution.  The photometric redshift calculated for the IR part only is equal to 3.28~$\pm$~0.43, while photometric redshift for the  UV--NIR part of the spectrum is 0.49. Observed fluxes are plotted with open blue squares. 
        Filled red circles correspond to the model fluxes. 
        The final model is plotted as a solid black line. 
        The remaining three lines correspond to the stellar, dust, and AGN components. 
        The relative residual fluxes, calculated as (observed flux - best model flux)/observed flux, are plotted at the bottom of each spectrum. } 
        \label{fig:SED_examp}
\end{figure*}

\subsection{Quality of the SED fits}
\label{sec:reduced_chi2}

The standard global quality of the fitted SEDs is quantified with the reduced value of $\chi^2$ ($\rm \chi^2$ divided by the number of data, hereafter: $\chi^2_r$) of the best model; it is not purely reduced $\chi{^2}$ regarding the statistical definition as the exact number of free parameters for each galaxy is unknown\footnote{the number of degrees of freedom can only be estimated for linear models; concerning nonlinear models, the number of degrees of freedom is absolutely nontrivial and whether or not it can be calculated properly is questionable; see e.g. \citep{Andrae:2010}  or \cite{Chevallard:2016}}. 
The minimum $\chi^2_r$ value for the galaxy still  points out the best model from the grid of all possible models created with the input parameters, but due to the varied number of observed fluxes  and unknown number of free parameters the  $\chi^2_r$ criteria cannot be use to remove galaxies
with unreliable fits from our sample.
Instead of $\chi^2_{r} $, we make use of the estimation of  physical properties ($\rm L_{dust}$ and $\rm M_{star}$) to select possible outliers  for the SED fitting procedure, and especially, galaxies which do not preserve the energy budget. 

To calculate the efficiency of using our method to select the ouliers, we choose galaxies for which we can estimate the $\rm L_{dust}$ based on the FIR measurements (the coverage of measurements, Fig.~\ref{fig:filters}, and quality of the optical data allow us to calculate stellar masses for all of the galaxies, and we do not need to make an additional cut for UV--OPT data). 
To obtain the most reliable sample of galaxies to test our method based on physical properties of galaxies we select objects with at least two measurements for FIR data with signal-to-noise rations (S/N)$\geqslant$3 (24~706 galaxies in total, $\sim$50\% of the ELAIS~N1 catalogue, hereafter: ELAIS~N1~3S/N).  

We run CIGALE two more times: (1) for optical measurements only, to estimate the classical stellar mass based on UV--OPT measurements and photometric redshifts only (hereafter: $\rm M_{star,stellar\mbox{ }SED}$), and (2) for FIR data only, to calculate dust luminosity (hereafter: $\rm L_{dust,IR\mbox{ }SED}$). 
In the following steps we compare $\rm L_{dust}$ and $\rm M_{star}$ values obtained from full UV-FIR SED fitting with  $\rm L_{dust,IR\mbox{ }SED}$ and $\rm M_{star,stellar\mbox{ }SED}$.

We then select galaxies with dust luminosity and/or stellar masses that are inconsistent  with the estimated based on the total SED fitting.
We use two criteria to find those objects:

\begin{itemize}
        \item Criterion 1: $\rm L_{dust} $ inconsistent (within 3$\sigma$ level) with  $\rm L_{dust, IR\mbox{ }SED} $, Fig.~\ref{fig:3sigma}a.
        \item Criterion 2:  $\rm M_{star} $ inconsistent (within 3$\sigma$ level) with  $\rm M_{star, stellar\mbox{ }SED} $, Fig.~\ref{fig:3sigma}b.
\end{itemize}

Based on the combination of these two criteria we select three different groups of objects: 

\begin{enumerate}[(i)]
        \item Criterion 1 not criterion 2: energy budget issue; we find 1~748 (7.08\% of the sample) galaxies with inconsistent $\rm L_{dust} $ and   $\rm L_{dust, IR\mbox{ }SED} $ and  consistent estimation of stellar mass (full magenta stars in Fig.~\ref{fig:3sigma}a) and we refer to them as galaxies with possible energy budget issues. Some of the sources  are possibly very deeply obscured galaxies (i.e. galaxy in Fig.~\ref{fig:SED_examp}a). After visual inspection we find that $\sim$ 25\% of them also show possible photometric redshift problems (incorrect matching between optical and IR counterparts).
        
        \item Criterion 2 not criterion 1: problem with matching optical catalogues; we find  353 objects  (1.43\% of the sample)  with inconsistent $\rm M_{star}$ estimation and  at the same time consistent estimation of dust luminosity (objects marked as yellow triangles in Fig.~\ref{fig:3sigma}b). Those objects are expected to have  problems with matching between catalogues\footnote{often with stars -- we find that 60\% of those objects have the nearest Gaia star source from GAIA DR1 \citep{Gaia2016} between 0.6 and 2 arcsec distance,  or that the stellarity parameter is larger than 0.5} or there is a problem with calibration between different optical data-sets used for the SED fitting  (Fig.~\ref{fig:SED_examp}b).
        \item Criteria 1 and 2:  268 objects (1.08\% of the sample) with inconsistency between runs in both stellar mass and dust luminosity (criteria 1 and 2, red open squares in Figs.~\ref{fig:3sigma}a and~b).
        The sub-sample is a mixture of stars, incorrect matching between stellar and IR catalogues, possible incorrect photometric redshifts and very small S/N for IR measurements. 
\end{enumerate}

In total, based on the physical analysis, we find only 2~369  (10\% of the ELAIS~N1~3S/N sample) peculiar objects (or catalogues  mismatches) that should not be included in the further physical analysis.

\subsection{Two $\chi^2$s criteria}
\label{sec:twochi2s}

\begin{figure}[h!]
        \begin{center}
                \includegraphics[width=0.5\textwidth]{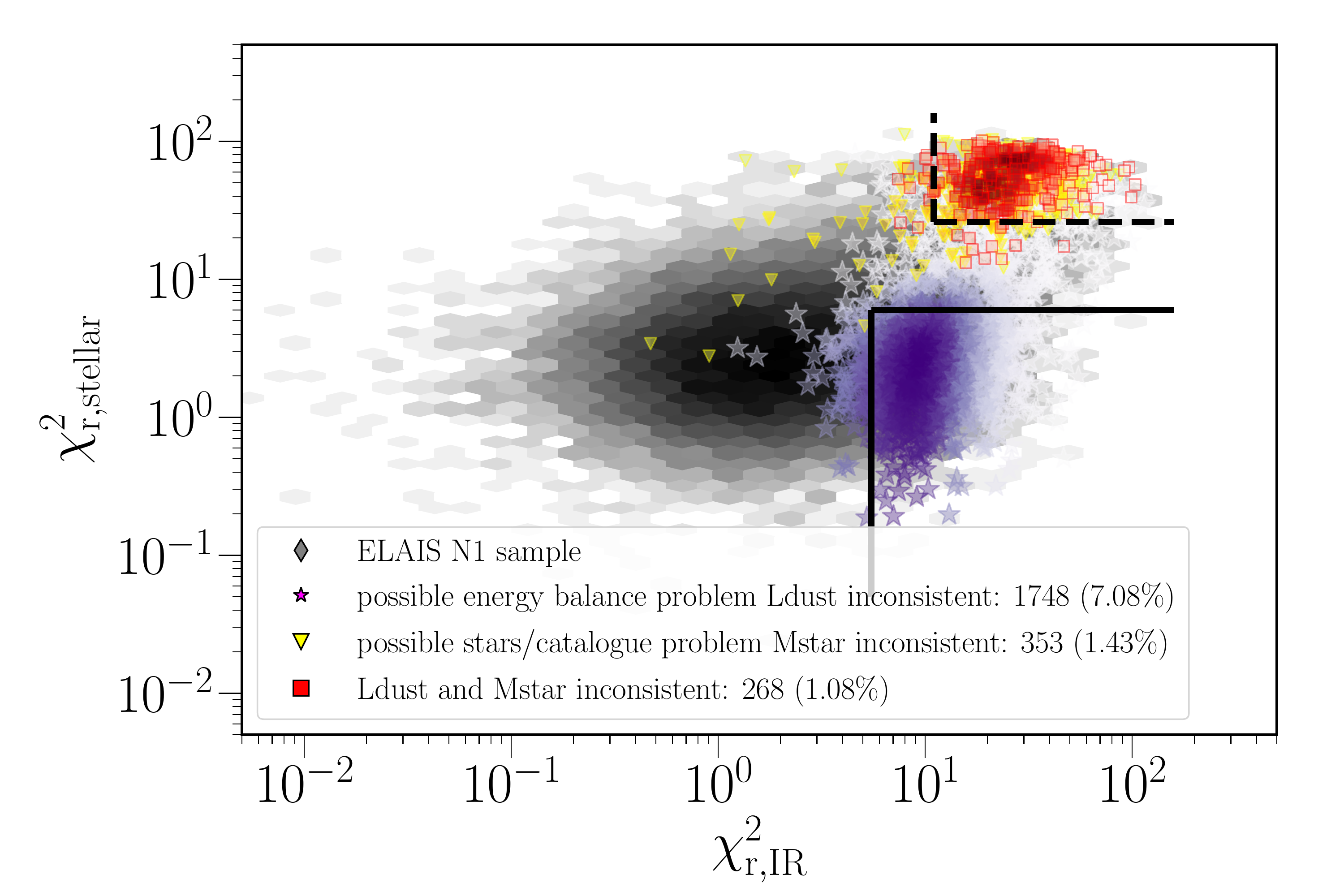}
        \end{center}
        \caption{Scatter between both $\chi^2$s. Grey-filled hexagons correspond to the  full ELAIS~N1 sample, magenta stars represent the sample of galaxies with possible energy budget problems, yellow triangles correspond to the galaxies with a possible problem with optical survey matching or stars, and red open squares correspond to objects showing inconsistency for both physical parameters used in our analysis. } \label{fig:chi2_scatter}
\end{figure}   

To try to remove the objects with inconsistent measurements of  $\rm L_{dust}$ (possible energy budget issue) and/or $\rm M_{star}$ (possible problem with matching optical catalogues) labelled  in Sect.~\ref{sec:reduced_chi2}, we use a combination of different  quantities describing the quality  of the fit. 
Our selection of galaxies can be correlated with $\rm \chi^2$s calculated for the stellar (hereafter: $\rm \chi^2_{r,stellar}$) and the dust (hereafter: $\rm \chi^2_{r,IR}$)  parts of the spectra, separately\footnote{We defined the wavelength ranges for  $\rm \chi^2_{r,stellar}$ and $\rm \chi^2_{r,IR}$ as $\leq$8~$\mu$m and $>$8~$\mu$m in rest frame, respectively, and calculate the values directly based on the best model.}. 

Figure~\ref{fig:chi2_scatter} shows the scatter between both $\rm \chi^2$s with groups of outliers found based on the analysis of physical parameters.    
We define two rejection criteria used to isolate at least 80\% of rejected objects as:

\begin{equation}
\begin{aligned} \label{eq:EB}
\rm \chi^2_{r,IR}\geqslant 5.5 \wedge \chi^2_{r,stellar}\leqslant 6,  \text{  black solid lines in Fig.~\ref{fig:chi2_scatter}}, \\
 \chi^2_{r,IR}\geqslant 11 \wedge \chi^2_{r,stellar}\geqslant26, \text{  black dashed lines in Fig.~\ref{fig:chi2_scatter}.} 
\end{aligned}
\end{equation}
This simple cut based on the values of $\rm \chi^2_{r,IR}$ and $\rm \chi^2_{r,stellar}$ allows us to reject:
\begin{itemize}
\item 81\% of objects with a possible problem with energy balance (magenta stars in Figs.~\ref{fig:3sigma}a and ~\ref{fig:chi2_scatter}, referred to as group (i) in Sect.~\ref{sec:reduced_chi2},
\item 85\% of objects with a possible incorrect match of optical survey or stars (yellow triangles in Figs.~\ref{fig:3sigma}b and ~\ref{fig:chi2_scatter}, group (ii) in Sect.~\ref{sec:reduced_chi2}),
\item and 92\% of galaxies with both $\rm M_{star}$ and $\rm L_{dust}$ parameters inconsistent with single runs for stellar and IR parts, respectively (open red squares in Figs.~\ref{fig:3sigma}a,~b and~\ref{fig:chi2_scatter}, group (iii) in Sect.~\ref{sec:reduced_chi2}).
\end{itemize}

We stress that the rejection criteria in Eq.~\ref{eq:EB} are very conservative in order to avoid excluding too many sources;  the full sample with all calculated $\rm \chi^2$s will be released to allow the user to make their own quality cuts.  
Values used for Eq.~\ref{eq:EB} are valid for ELAIS~N1 field, and the thresholds for other HELP fields should be revised. 

Our two $\chi^2$s criteria (Eq.~\ref{eq:EB}) remove an additional 2~634 galaxies (5.25\% of the sample) lying below magenta stars in Fig.~\ref{fig:chi2_scatter}.  
We check what kind of objects we reject. 
We suspect that some of the sources have problems with matching between  optical and IR data and that the rejected sample includes possible incorrect photometric redshift, as in, for example, lensed galaxies, which are a good example of objects with energy budget issues. 

To check whether or not the  mismatch between optical and FIR data is at the basis of the rejection, we calculate $\rm z_{phot,IR}$ - photometric redshifts based on the PACS and SPIRE data only. 
We make a standard check for photometric redshift outliers \citep{Ilbert:2006}: $ \rm \delta(redshift-z_{phot,IR})/(1+redshift)>N$, where redshift corresponds to the photometric redshift calculated based on UV--OPT data. 
In the literature, comparisons of photometric and spectroscopic data use a value of N equal to 0.15, but this could be too small for comparison between two photometric values.  
We find that using double this  value (N=0.30, more suitable for less robust estimation) criteria  select 78\% of rejected objects. 
Requiring at least three IR measurements with S/N$\geqslant$3 for ELAIS~N1 catalogues to obtain secure  $\rm z_{phot,IR}$ estimates  reduces the ELAIS~N1~3S/N sub-sample to 1~059 objects only.  
We also find that for this secure sample, only  10\% of objects  not rejected by the $\rm \chi^2$s, and 99\% of galaxies selected by Eqs.~\ref{eq:EB}, have a possible problem with photometric redshifts. 
This suggest that the majority of rejected galaxies are incorrect matches between optical and FIR data or peculiar galaxies, as for example lensed galaxies.   
To summarise, the contamination of rejected objects is as low as  $\sim$12\%. 

We checked all objects rejected by Eqs.~\ref{eq:EB} (5~291 galaxies in total) and found more than 300 possible lensed candidates inside the criteria describing galaxies with possible energy budget problem (marked in Fig.~\ref{fig:chi2_scatter} as black solid line). 
Based on visual inspection we find that the difference between photometric redshifts estimated based on the UV--OPT  and IR measurements for all of them is higher than 0.63 (twice the mean uncertainties obtained for the $\rm z_{phot,IR}$). 
An example of a possible lensed candidate is shown in  Fig.~\ref{fig:SED_examp}c.

Using two $\chi^2$s to select peculiar objects is not a new technique (but used for the first time in CIGALE tool).  
For example, \cite{MRR2014} used the quality of the fit of the IR part of the galaxy spectra to select possible lensing candidates out of 967 SPIRE sources in the HerMES Lockman survey and found 109 candidates. 
We checked if our method to select lensed candidates (object rejected by Eqs.~\ref{eq:EB}  $\wedge$ $\rm |redshift-z_{phot,IR}|>0.63 $,  the nearest Gaia star is further than 0.6 arcsec (we used flag  $gaia\_flag==0$  from the HELP catalogue) and stellarity parameter$<$0.3) is in agreement with the colour-redshift criteria provided by \cite{MRR2014} for lensed candidates at redshift 0.15--0.95. 
We find that 70\% of our possible lensed galaxies at this redshift range fulfills conditions given by \cite{MRR2014}.    
This proves that this method  provides  very useful information about the disagreement between the stellar and dust parts of the galaxy spectrum. 
This information is even more useful for SED fitting, where we are using code that preserves the energy balance. 

We also check to see if the galaxies lying between the $\rm \chi^2_{r,stellar}$ rejection criteria  (galaxies defined by 5.5 $\rm < \chi^2_{r,sterllar} <$ 26 and $\rm \chi^2_{r,IR}>$11) have special properties. 
We find that the majority of them are characterised by very small error bars for UV--NIR measurements and with much higher flux uncertainties for \textit{Herschel} fluxes. 
This implies imprecise SED fitting, but at the same time, both $\rm M_{star}$ and $\rm L_{dust}$ , calculated separately for UV--IR and IR data, have similar values (below 3$\sigma$ threshold, with high uncertainties coming from PDF analysis).  
Nonetheless, we have no solid basis to remove those objects from our analysis. 

To provide general properties of the  ELAIS~N1 sample we  removed 7~533 galaxies in total based on the two $\chi^2$s cuts (Eqs.~\ref{eq:EB}). 
We define the remaining 42~453 galaxies (84\% of the original ELAIS~N1 sample) as a clean sample for the physical analysis.
The redshift distribution of the final sample used for SED fitting is presented in Fig.~\ref{fig:redshift} (the red hatched histogram).

We stress that  $\rm \chi^2_r$, $\rm \chi^{2}_{r,stellar}$, and $\rm \chi^{2}_{r,IR}$ are a part of deliverables and each user is free to choose different criteria (or to not use them at all). 
For each field we provide the main physical parameters without any subjective flags describing the quality of the fit. 
Both, the data and the SED fitting results  can be downloaded at \url{http://hedam.lam.fr/HELP/dataproducts/dmu28/dmu28_ELAIS-N1/data/zphot/} and also accessed with Virtual Observatory standard protocols at \url{https://herschel-vos.phys.sussex.ac.uk/}. 

\renewcommand{\arraystretch}{1.15}
\begin{table*}[]
        \begin{center} 
        \caption[] {The main physical properties of the sample of  42~453 galaxies presented  in  total IR luminosity bins. The first column corresponds to the galaxy type according to their total $\rm log(L_{dust})$ value, the second shows the number of galaxies (first row), and the total percentage of the sample (second row), column (3) presents the number of galaxies with an AGN contribution higher than 20\% (first row -- total number, second row -- percentage value of galaxiwhiches with AGN contribution within the given IR luminosity bin). Median values for redshift, stellar mass, SFR and specific SFR are given in columns (4) - (7). The errors are calculated as median absolute deviation.} 
        \label{tab:statystyka}
        \begin{tabular}{l|c|c|c|c|c|c} \hline \hline  
        \multirow{3}{*}{ } & \multirow{3}{*}{N$_{gal}$} & \multirow{3}{*}{N$_{AGN}$} &   \multirow{3}{*}{redshift} & \multirow{3}{*}{$\rm log(M_{star})$} & \multirow{3}{*}{$\rm log(SFR)$} & \multirow{3}{*}{$\rm log(sSFR)$}  \\
        & & & & & & \\
        & (\% ) & (\%)  &  & $\rm [M_{\odot}]$ & $\rm [M_{\odot}yr{^-1}]$ & $\rm [yr^{-1}]$   \\ \hline 
        (1) & (2) & (3) & (4) & (5) & (6) & (7) \\ \hline
        \multirow{2}{*}{{\footnotesize $\rm log(L_{dust})<11$}} & 6~974 &  268 & \multirow{2}{*}{0.42 $\pm$ 0.15} &       \multirow{2}{*}{10.12 $\pm$ 0.34}  &        \multirow{2}{*}{0.64 $\pm$ 0.27} &\multirow{2}{*}{-9.70 $\pm$ 0.16}  \\ 
        & 16.43\% & 3.86\% & &  &   &  \\ \hline
        \multirow{2}{*}{LIRGs \mbox{ } {\footnotesize $\rm 11\leq log(L_{dust})<12$}} & 23~214 &  724 & \multirow{2}{*}{0.96 $\pm$ 0.27} & \multirow{2}{*}{10.73 $\pm$ 0.24} & \multirow{2}{*}{1.59 $\pm$ 0.23 } & \multirow{2}{*}{-9.22 $\pm$ 0.31} \\
        & 54.68\% & 3.11\% &  &  &  &  \\ \hline
        \multirow{2}{*}{ULIRGs\mbox{ } {\footnotesize $\rm 12\leq log(L_{dust})<13$}} & 11~947&  705 & \multirow{2}{*}{1.89 $\pm$ 0.32}& \multirow{2}{*}{11.10 $\pm$ 0.22} & \multirow{2}{*}{2.30 $\pm$ 0.17} & \multirow{2}{*}{-8.8{5} $\pm$ 0.21}  \\
        & 2{8.14}\% & 5.{90}\% &  &  &  &  \\ \hline
        \multirow{2}{*}{HLIRGs \mbox{ } {\footnotesize $\rm log(L_{dust})\geq13$}} & {318} &  {65} & \multirow{2}{*}{4.9{2} $\pm$ 0.3{4}} & \multirow{2}{*}{11.4{9} $\pm$ 0.3{4}} & \multirow{2}{*}{3.5{2} $\pm$ 0.17} & \multirow{2}{*}{-8.01 $\pm$ 0.5{3}} \\
        & 0.7{5}\% & 20.4{4}\% &  &  &  &  \\ \hline
        \end{tabular} 
\end{center}
\end{table*}

\section{General properties of the ELAIS~N1 sample}
\label{sec:General_properties}

\subsection{Dust luminosity distribution for ELAIS~N1}
        
The majority of our sample consists of  (LIRGs) and ultra luminous infrared galaxies (ULIRGs), characterised by $\rm 11\leq log(L_{dust}/L_{\odot})<12$, and   $\rm 12\leq log(L_{dust}/L_{\odot})<13$, respectively. 
We have found 2{3~214} LIRGs (5{4.68}\% of the sample), 11~{947} ULIRGs (2{8.14}\%), and 318 (0.7{5}\%) hyper LIRGs (HLIRGs) ($\rm log(L_{dust}/L_{\odot})\geq13$). 
The main statistical properties of LIRGs, ULIRGs, and HLIRGs, as well as the galaxies characterised by $\rm log(L_{dust}/L_{\odot})<11$ from the ELAIS~N1 field, can be found in Table~\ref{tab:statystyka}.
All delivered parameters (dust luminosity, stellar mass, SFR,  $\chi^2_r$, $\rm \chi^2_{r,stellar}$, $\rm \chi^2_{r,IR}$, and calculated thresholds) will be published while the HELP database is being completed (end of 2018).

\subsection{AGN contribution}
        
{\renewcommand{\arraystretch}{1.2}
\begin{table}[h!]
        \begin{center} 
                \caption[]{NIR criteria for an AGN selection. SED AGNs  correspond to the number of AGNs found by SED fitting and fulfilling  given criterion; numbers in brackets represent the corresponding percentage taking into account the  full SED AGN sample of 5{80} objects with IRAC data. }
                \label{tab:AGN_criteria}
                \begin{tabular}{l | c } \hline  
                        Criterion & SED AGNs    \\ \hline 
                        \cite{Stern2005AGNselection} &  {393 (67.7\%)} \\
                        \cite{Donley2012AGNselection} & {339 (58.4\%)} \\
                        \cite{Lacy2007AGNselection} &   {435 (75.0\%)} \\
                        \cite{Lacy2013AGNselection} &   {485 (83.6\%)} \\ \hline
                \end{tabular}
        \end{center}
\end{table}
}

\begin{figure*}[h!]
        \begin{center}
                \includegraphics[width=0.99\textwidth,clip]{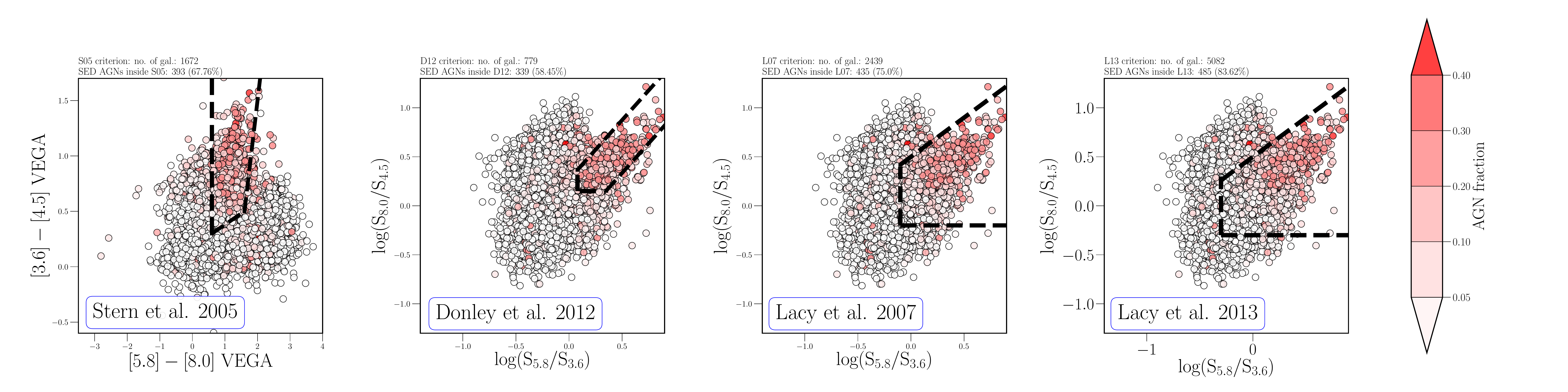}    
        \end{center}
        \caption{ \cite{Stern2005AGNselection}, \cite{Donley2012AGNselection}, \cite{Lacy2007AGNselection} and \cite{Lacy2013AGNselection} MIR selection of AGNs. Black dashed lines correspond to the criteria. 546 galaxies from our sample for which the $AGN_{frac}>=0.2$ are marked in red, according to the AGN fraction.}
        \label{fig:AGNcriteria}
\end{figure*}

As it was shown in, for example,  \cite{Ciesla15}, \cite{Salim2016} and  \cite{Leja2018},  that not taking into account an AGN component when performing broad band SED fitting of AGN host galaxies results in substantial biases in their derived parameters.

The AGN fraction in CIGALE is calculated as the AGN contribution to dust luminosity obtained by SED fitting based on the bayesian analysis.
We find 1~{762}   galaxies (2{68} galaxies with $\rm log(L_{dust})<11$, {724} LIRGs, {705} ULIRGs and {65} HLIRG) with a significant ($\geqq$20\%)  AGN fraction.
We  refer to that sample as SED AGNs.
        
We performed a run of the SED fitting with the same parameters for physical models but without the AGN component to check how the stellar mass and the SFR differ between two runs. 
We find that  $\rm log(M_{star})$ is generally well recovered, even for galaxies with a large AGN contribution. 
Galaxies with a significant AGN contribution have higher SFR  when calculating without the AGN component, and  the difference can be as high as 0.5 dex. 
The result of the test is described in Appendix~\ref{app:with_without_test}.   
We conclude that the use of an AGN component is necessary to run CIGALE's SED fitting for HELP. 
        
We also check whether our method is in agreement with various criteria based on the IRAC colour selections. 
We select {9~196}  ELAIS~N1 galaxies (21.{66}\% of the final sample) which have measurements at all four IRAC bands. 
For 5{80} of them, we find an AGN component according to the CIGALE SED fitting. 
We focus on four different criteria for AGN selection based on the NIR data:  \cite{Stern2005AGNselection,Donley2012AGNselection,Lacy2007AGNselection,Lacy2013AGNselection}. 
The majority of the 5{80} SED AGNs meet the listed criteria (see Table~\ref{tab:AGN_criteria}).  
Figure~\ref{fig:AGNcriteria} shows criteria listed above (areas marked with black dashed lines) and galaxies with a significant AGN contribution according to the SED fitting (marked as red points). 
        
We conclude that the usage of the AGN module \citep{fritz06} for all ELAIS~N1 galaxies gives us a reasonable (according to NIR critera) sample of AGNs. 
For example,  only $\sim$2{7}\% (on average) of SED AGNs lie outside the NIR criteria of  \cite{Stern2005AGNselection,Donley2012AGNselection,Lacy2007AGNselection},  and \cite{Lacy2013AGNselection}, meaning that the contamination of the SED AGNs sample is small.   
In total the ELAIS~N1 sample contains only  4.{1}5\%  of galaxies with a significant AGN contribution and it has no important influence for the main properties of the full sample. 
        
In the framework of the HELP project, we deliver the physical properties of galaxies ($\rm M_{star}$, SFR, $\rm L_{dust}$). 
The AGN identification is a by-product of our analysis. 
Tests presented above show that the inclusion of emission from dusty AGNs does not generate significant biases for estimated properties.

\section{Dust attenuation recipes}
\label{sec:Dust_att}
        \begin{figure}[]
        \begin{center}
                \includegraphics[width=0.41\textwidth,clip]{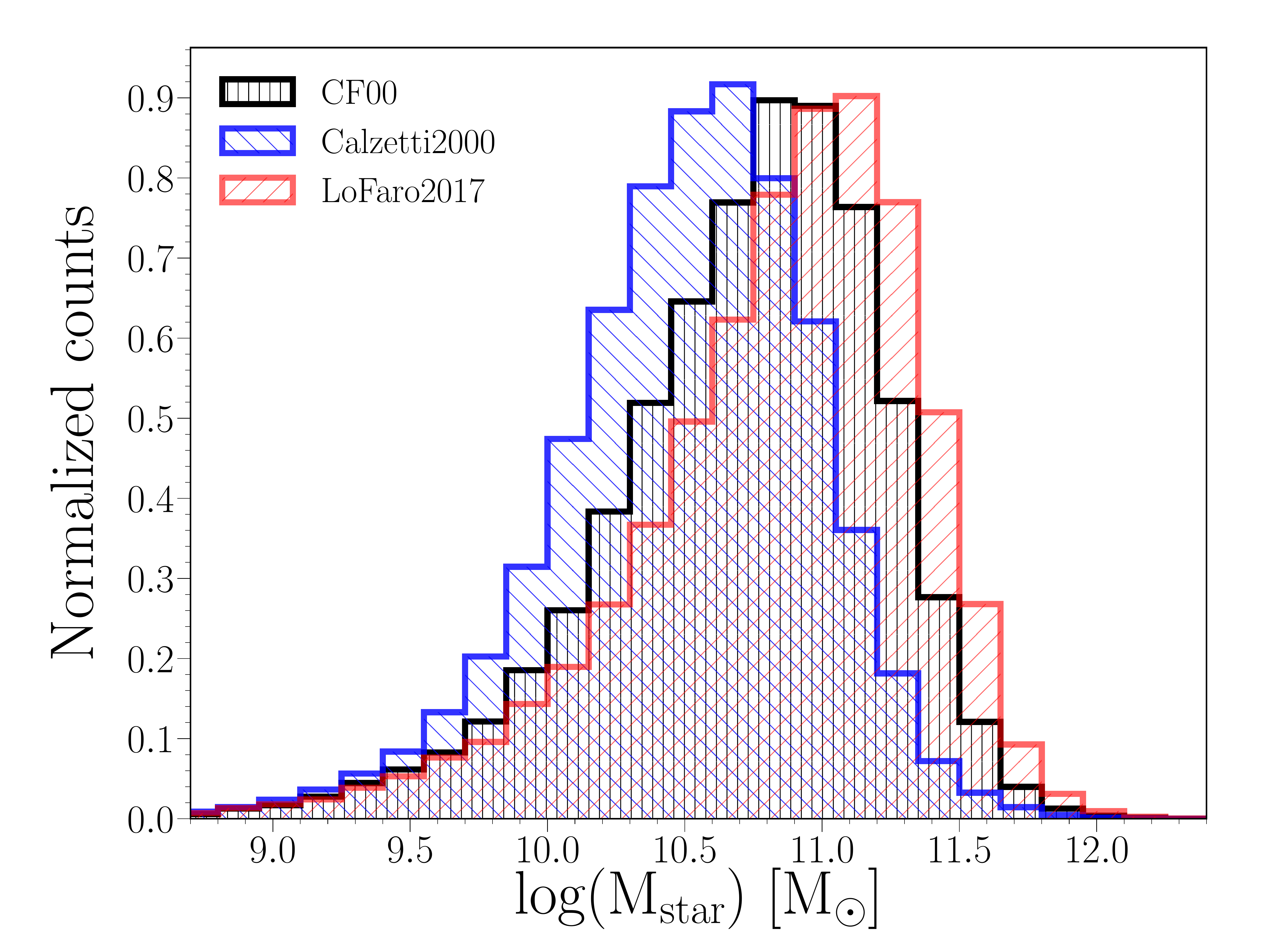}    
        \end{center}
        \caption{Comparison of stellar masses obtained with three different dust attenuation laws: \cite{CF00} -- black striped histogram, \cite{calzetti00} - blue right hatched histogram, and \cite{LoFaro2017} - red left hatched histogram.  }
        \label{fig:Mstar_histo_comparison}
\end{figure}

\begin{figure}[h!]
        \begin{center}
                \includegraphics[width=0.41\textwidth,clip]{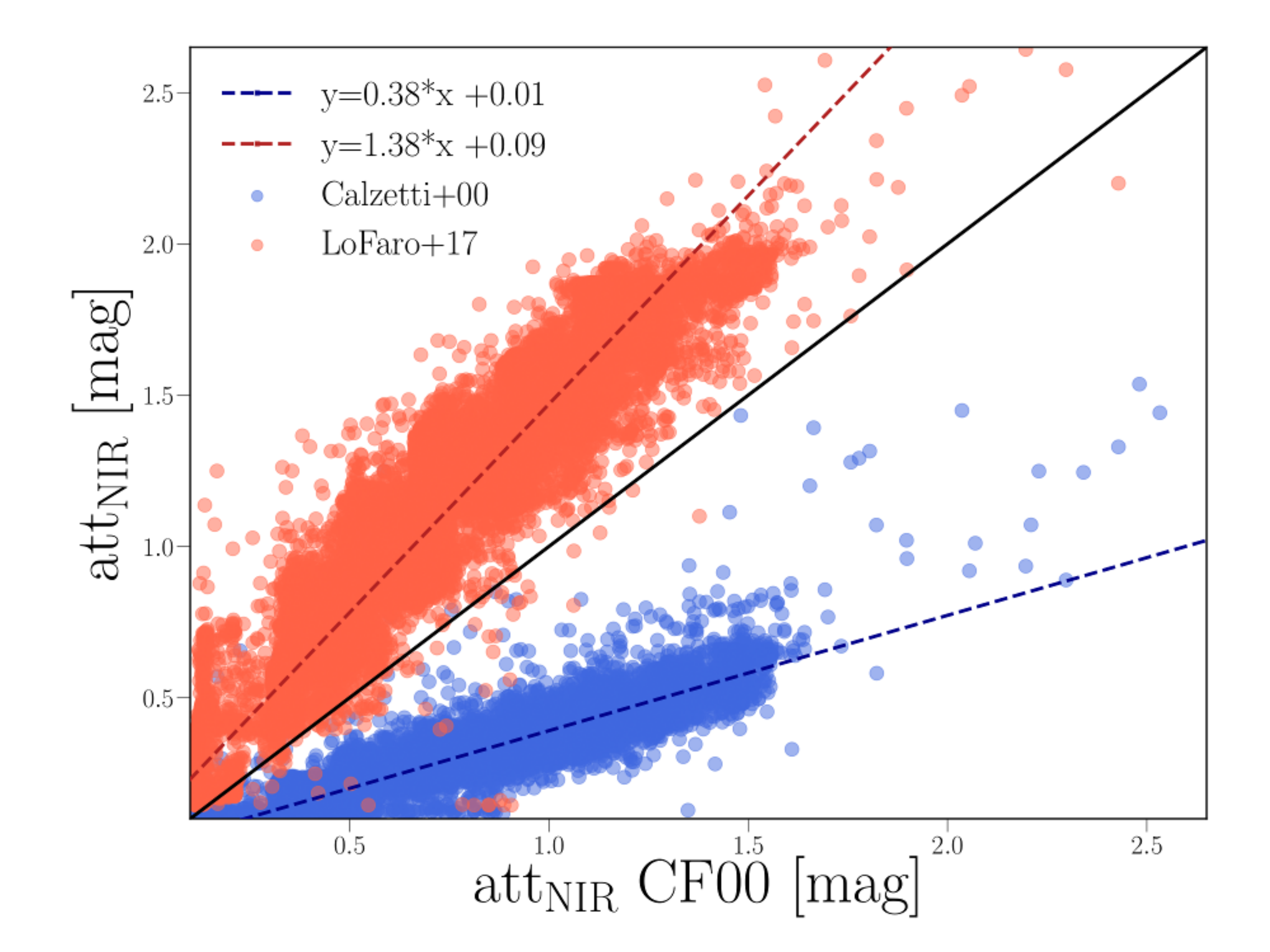}
        \end{center}
        \caption{Total dust attenuation in NIR filter estimated under the assumption of the \cite{CF00} dust attenuation curve with the Calzzetti dust attenuation law (full blue dots) and Lo~Faro model (full red dots). The black solid line corresponds to the 1:1 relation while blue and red dashed lines correspond to the relation between \cite{CF00} and \cite{calzetti00}, and \cite{CF00} and \cite{LoFaro2017} attenuation laws, respectively. }
        \label{fig:att_h}
\end{figure}

\begin{figure*}[ht!]
        \begin{center}
                \subfloat[$\rm log(L_{dust})$]
                {\includegraphics[width=0.26\textwidth]{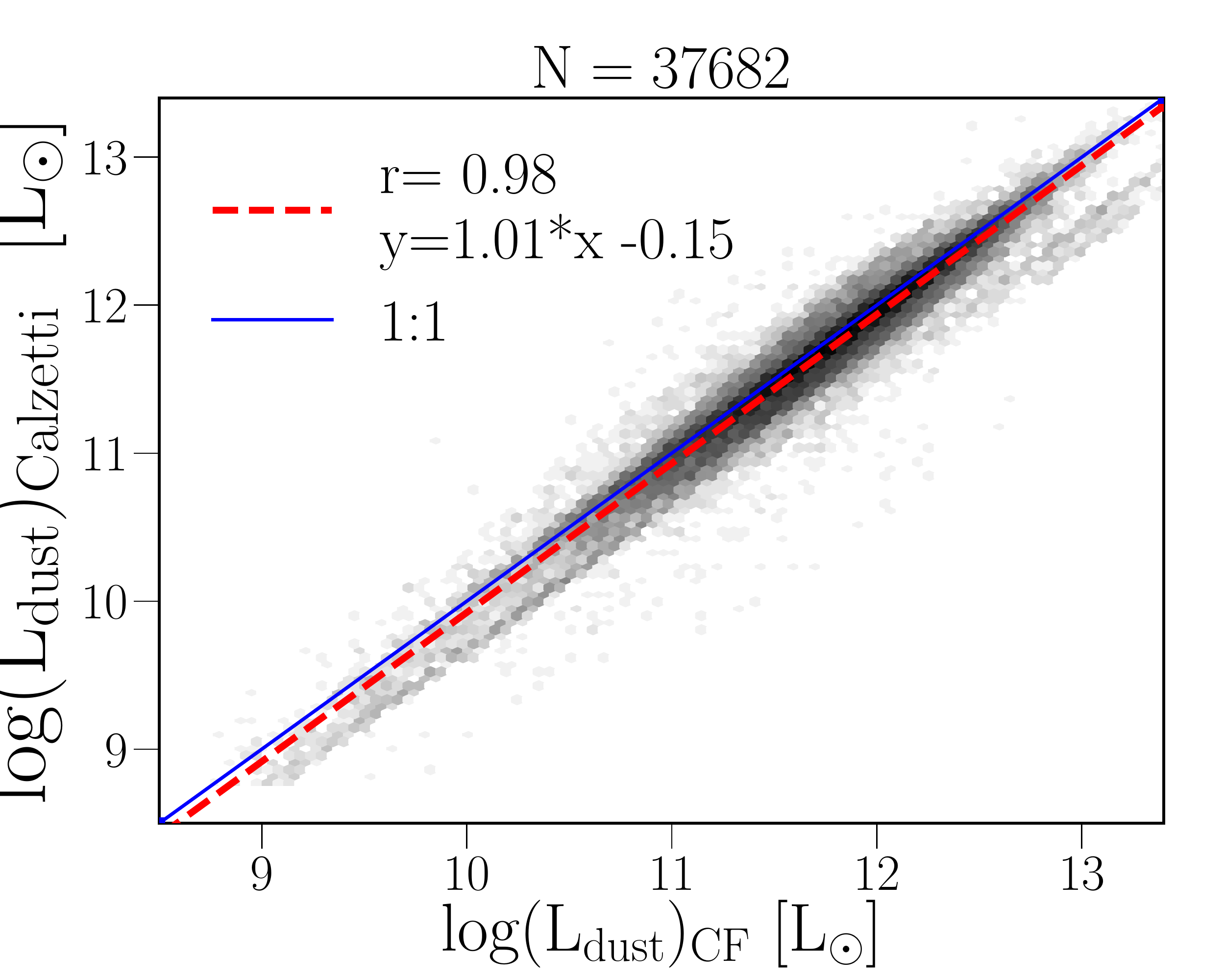}} 
                \subfloat[$\rm log(SFR)$]
                {\includegraphics[width=0.26\textwidth]{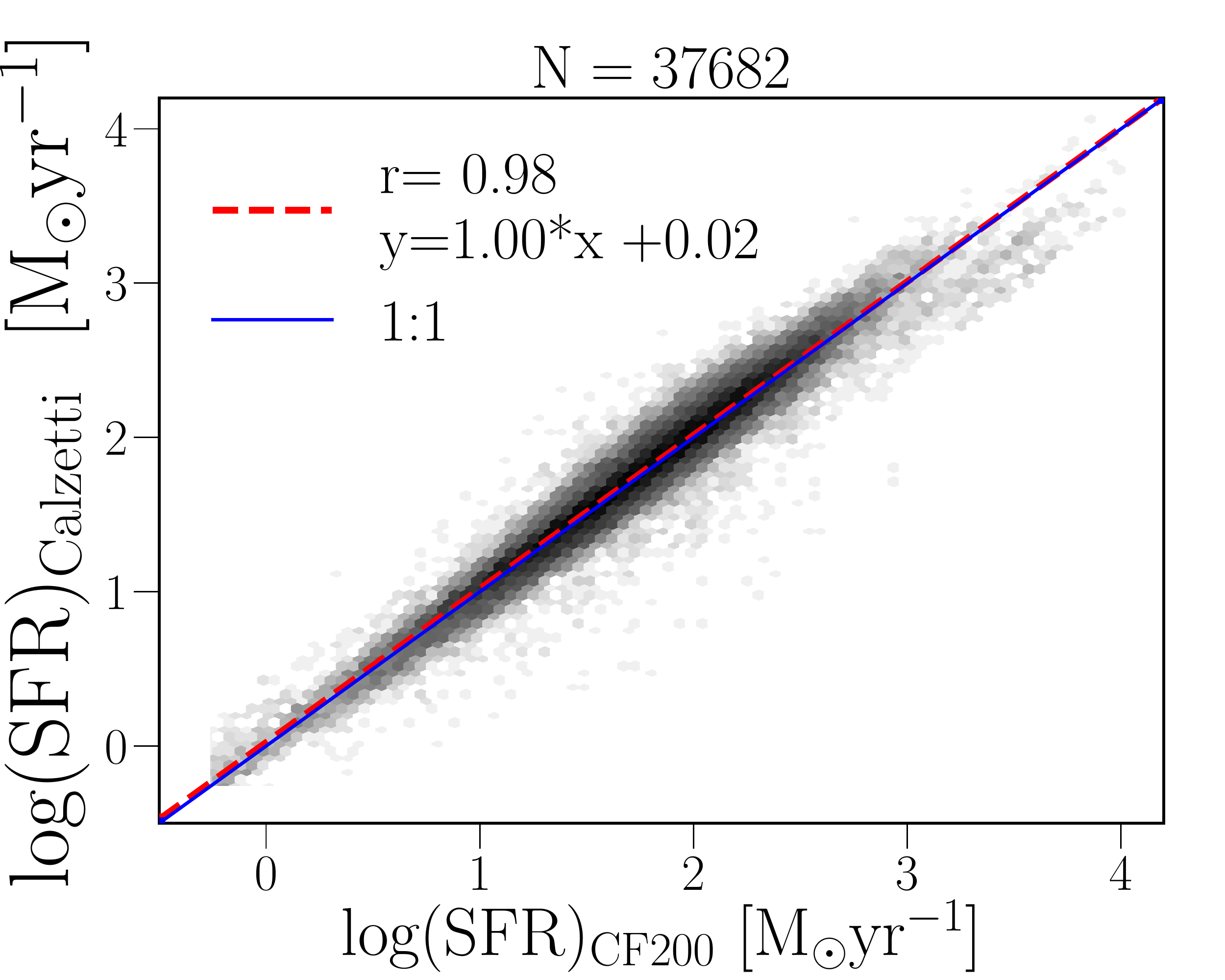}}
                \subfloat[ $\rm log(M_{star})$ ] {\includegraphics[width=0.26\textwidth]{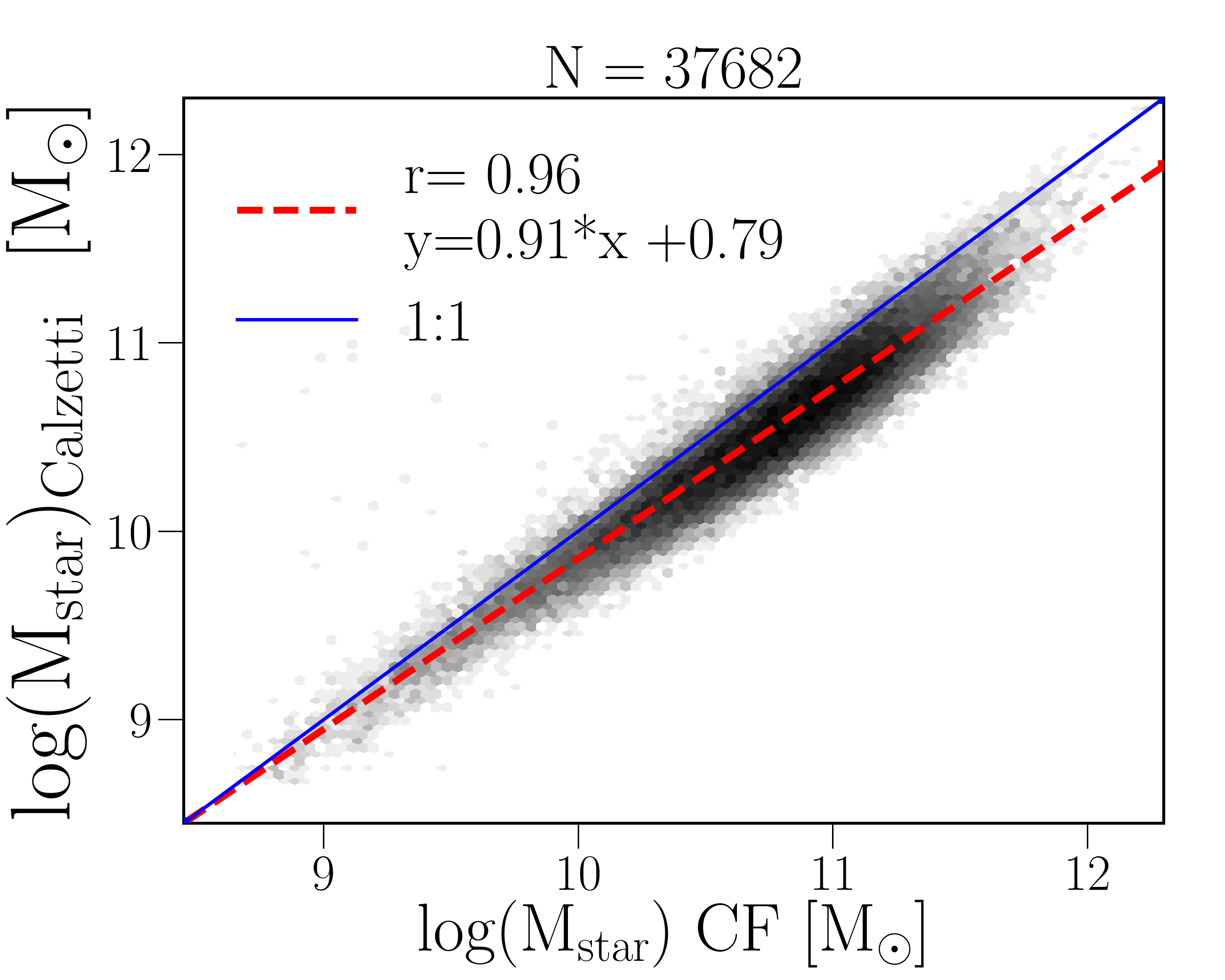}}
                
                \subfloat[$\rm log(L_{dust})$]
                {\includegraphics[width=0.26\textwidth]{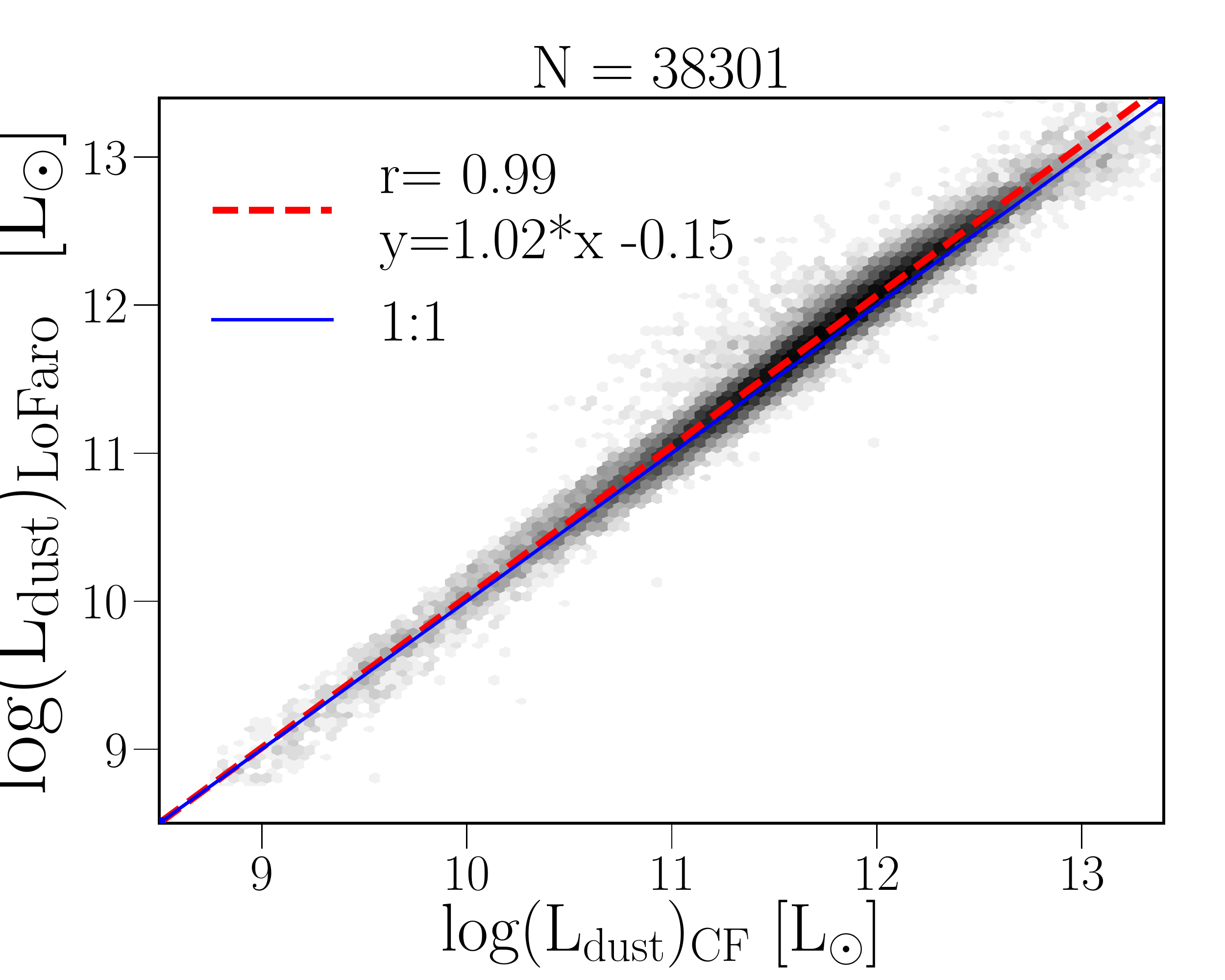}} 
                \subfloat[$\rm log(SFR)$]
                {\includegraphics[width=0.26\textwidth]{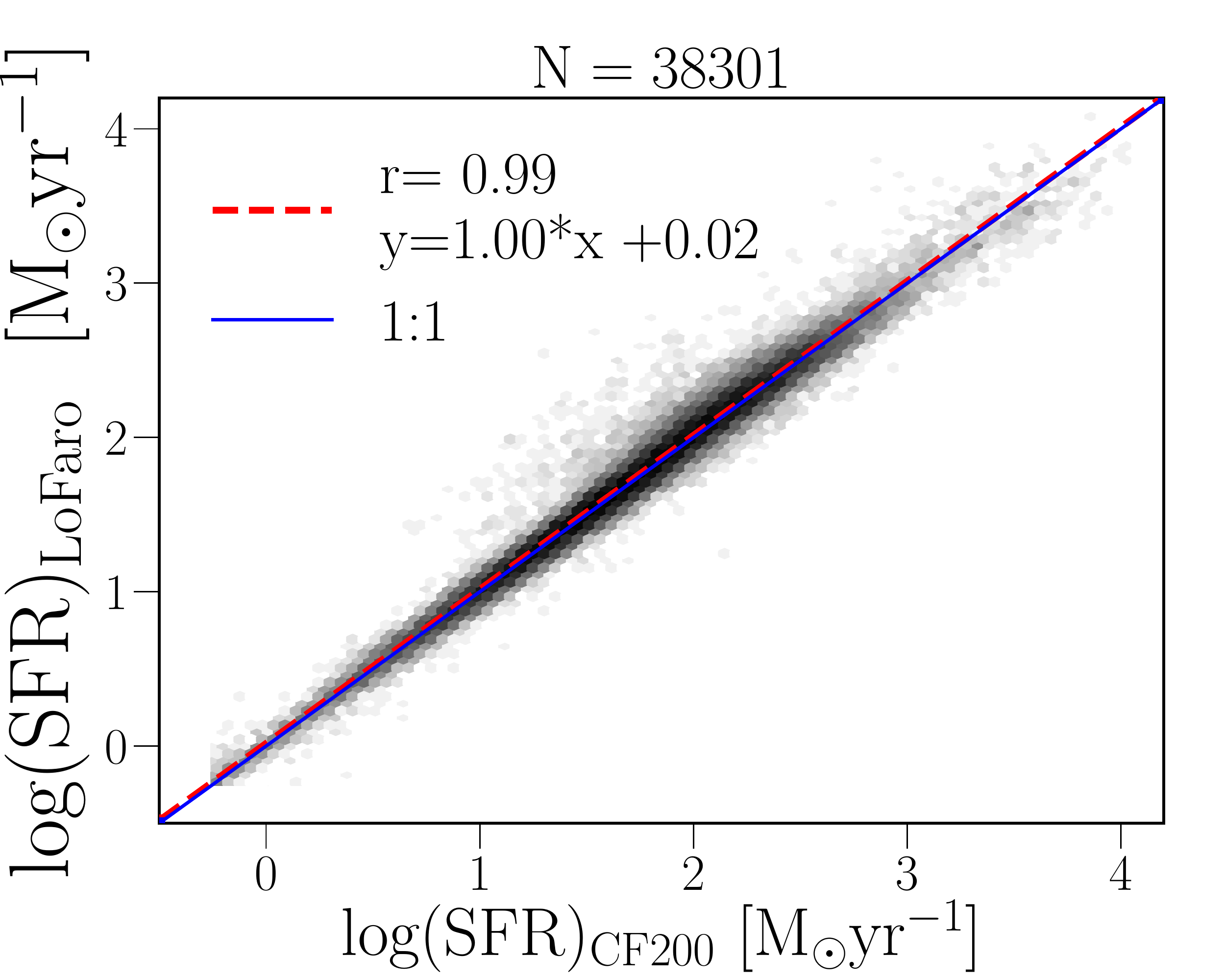}}
                \subfloat[ $\rm log(M_{star})$ ] {\includegraphics[width=0.26\textwidth]{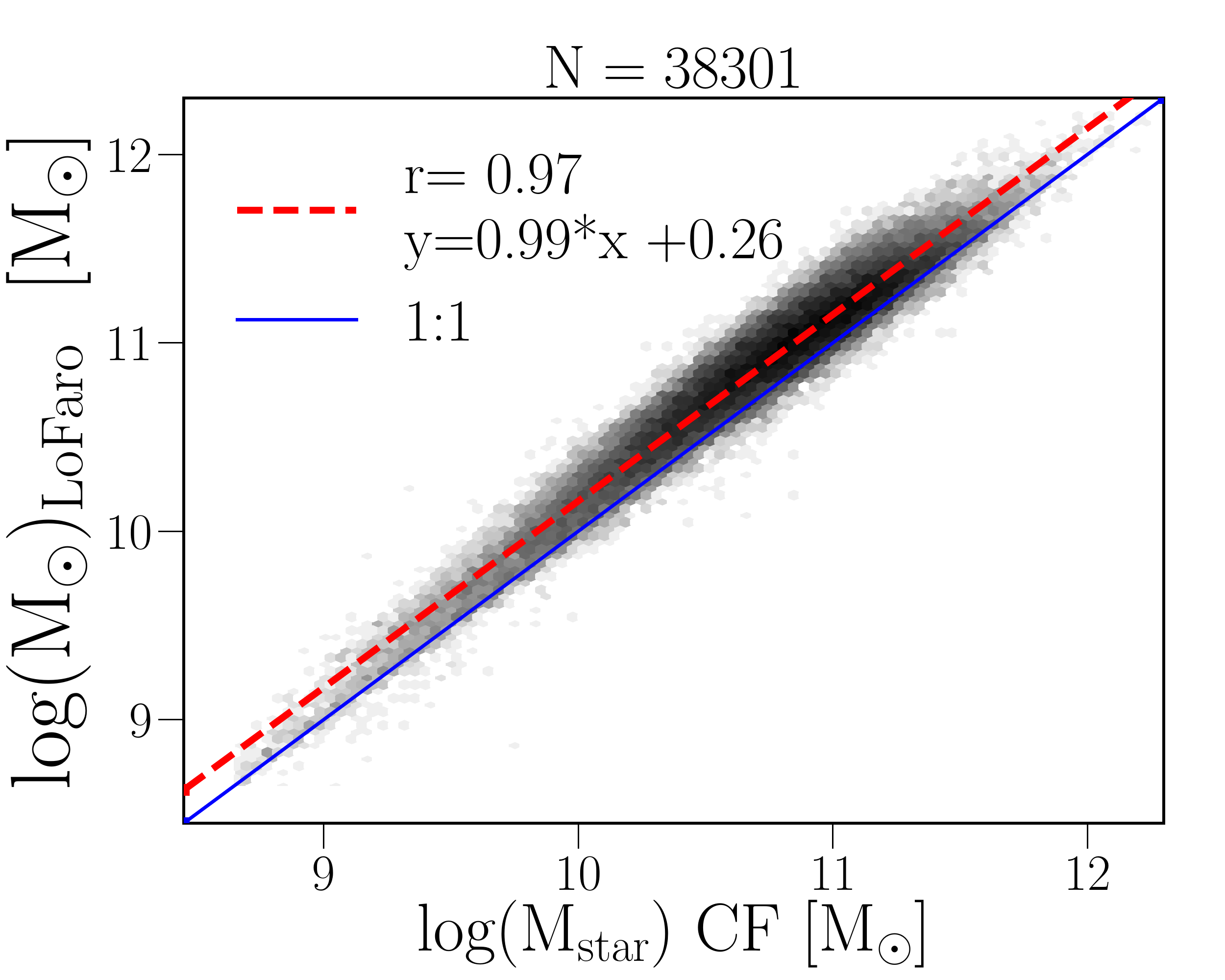}}
        \end{center}
        \caption{Comparison of main physical parameters ($\rm log(L_{dust})$, $\rm log(SFR)$, and  $\rm log(M_{star})$) for a sample of 37~682 galaxies fitted with the \citealp{calzetti00} attenuation law (y-axis), or the \citealp{CF00} attenuation law (x-axis) (\textit{upper panel}), and for a sample of 38~301 galaxies  fitted with the \citealp{LoFaro2017} attenuation law (y-axis), or the \citealp{CF00} attenuation law (x-axis) (\textit{bottom panel}).
                1:1 relation is marked with blue solid line, fitted relation - as red dashed line. Each panel includes Pearson product-moment correlation coefficient (\texttt{r}). } 
        \label{fig:Caltzetti_vs_CF}
\end{figure*}

Based on our sample we explore how physical parameters obtained by fitting SEDs with different dust attenuation models can be biased.
We apply three different attenuation laws to explore this subject. 
We perform two additional SED fitting runs: (1) the \cite{calzetti00} recipe that is widely used in the literature,
  and (2) the \cite{LoFaro2017} law obtained for IR bright galaxies.
The main change between those attenuation laws is the slope for the part of the spectrum with $\lambda>$5000~$\AA$  where the \citetalias{CF00} law is much flatter in both the UV and visible part of the spectrum than \cite{calzetti00} and a bit steeper than \cite{LoFaro2017}. 
Figure~1 in \cite{LoFaro2017} highlights the main differences between all three recipes for dust attenuation considered in this paper. 
        
We find, based on the quality of the fits, that 45\% of galaxies are fitted better with \citetalias{CF00} than the other two attenuation laws. 
The \cite{LoFaro2017} procedure works very well for 24\% of cases  and \cite{calzetti00} for 29\%. 
It was expected that \citetalias{CF00} and \cite{LoFaro2017} attenuation laws would work better for our sample as   \cite{LoFaro2017} has shown that the \cite{calzetti00} dust model cannot accurately reproduce the attenuation in a sample of ULIRGs at redshift$\sim$2 (around 30\% of our sample consists of dusty galaxies at redshift $\sim$2, and more than half of the objects have $\rm log(L_{dust})\geq$11.5~[$\rm L_{\odot}$]).
        
As was already shown in \cite{LoFaro2017} and \cite{Mitchell2013}, the influence of the attenuation curve on the derived stellar masses can be strong \citep[e.g. ][showed that for the most extreme cases the difference can reach a factor of 10, with the median value around 1.4]{Mitchell2013} compared to greyer attenuation curves with the standard \cite{calzetti00} recipe. 
A flatter attenuation curve at longer wavelengths results in larger stellar masses.     
We now compare these findings to our ELAIS-N1 sample. 
For both additional runs with \cite{calzetti00} and \cite{LoFaro2017} recipes, we used exactly the same parameters for the other  modules.  
All parameters used for SED fitting are listed in Table~\ref{tab:input}.  
        
For comparison we use only galaxies without possible outliers (we used Eqs.~\ref{eq:EB} presented is Sect.~\ref{sec:twochi2s} for runs with \cite{calzetti00} and \cite{LoFaro2017}  attenuation laws).  
In total we perform a comparison between \citetalias{CF00} and \cite{calzetti00} for 37~682 galaxies, and \citealp{CF00} and \cite{LoFaro2017} for 38~301 objects. 
We find an agreement between all three runs for $\rm L_{dust}$ and SFR (1:1 with slight scatter: see Fig.~\ref{fig:Caltzetti_vs_CF} for details). 
We conclude  that using either of  the attenuation laws of \citetalias{CF00}, \cite{calzetti00}, or \cite{LoFaro2017}  has no important impact on dust luminosity or SFR estimation.
Nevertheless, the choice of attenuation law has a substantial impact on calculation of stellar mass.

The right panels of Fig.~\ref{fig:Caltzetti_vs_CF} show the difference between $\rm log(M_{star})$ obtained with \citetalias{CF00} and  \cite{calzetti00} (upper panel) and \citetalias{CF00} and \cite{LoFaro2017} (bottom panel) laws. 
Here we can see a significant shift between both values, which becomes stronger for more massive galaxies for Calzetti versus CF00 attenuation laws. 
        
Figure~\ref{fig:Mstar_histo_comparison} shows $\rm log(M_{star})$ distributions obtained for runs with three different attenuation laws. 
It is clearly visible that stellar masses obtained with the \cite{calzetti00} law are on average lower than those from \citetalias{CF00} and \cite{LoFaro2017}. 
Derived mean values of $\rm log(M_{star})\mbox{ }[M_{\odot}]$ are equal to 10.50~$\pm$~0.47,  10.72~$\pm$~0.49, and 10.83~$\pm$~0.54, for \cite{calzetti00}, \cite{CF00}, and \cite{LoFaro2017}, respectively.

We find the relation between stellar masses obtained with different attenuation laws:
For \cite{calzetti00} and \citetalias{CF00}:
\begin{equation}
\label{eq:Mstar_full_SED_fitting}
\rm log(M_{star})_{Calzetti}=0.91 \times log(M_{star})_{CF00} + 0.79, 
\end{equation} 
and for \cite{LoFaro2017} and \citetalias{CF00}:
\begin{equation}
\rm log(M_{star})_{LoFaro}=0.99 \times log(M_{star})_{CF00} + 0.26. 
\end{equation}   

We note that we find the same relations excluding IR data from the SED fitting. 
Our analysis of the influence of the dust attenuation law for estimated  dust luminosity, SFR, and stellar mass based on the stellar emission only is presented in Appendix~\ref{app:attenuation_laws}.

As was also checked by \cite{LoFaro2017} for a sample of IR selected [U]LIRGs,  the total amount of attenuation for a young population (in UV ranges) is usually well constrained by different  attenuation laws, but it this is not the case in NIR wavelength range. 
We find that attenuation in the NIR range is not preserved between different attenuation laws, and is on average larger for \cite{LoFaro2017} and \cite{CF00} than for \cite{calzetti00} (Fig.~\ref{fig:att_h}). 
This relation is a direct  outcome of the shape of  adopted attenuation curves.
We find that the ratio between mean $\rm M_{star}$ obtained from \cite{calzetti00} and \citetalias{CF00} is equal to 0.98$\pm$0.14, and that the ratio between mean $\rm M_{star}$ obtained from \cite{LoFaro2017} and \citetalias{CF00} is 1.49$\pm$0.44. 
The mean ratio between stellar mass calculated based on the NIR band magnitude from mass to luminosity relation gives very similar values: 0.98$\pm$0.01 and 1.41$\pm$0.24 for Calzetti/CF00 and LoFaro/CF00, respectively. 
We summarise that the shape of the attenuation laws in the NIR band is responsible for inconsistency in  stellar masses.   

\begin{figure*}[]
        \begin{center}
                \includegraphics[width=0.95\textwidth,clip]{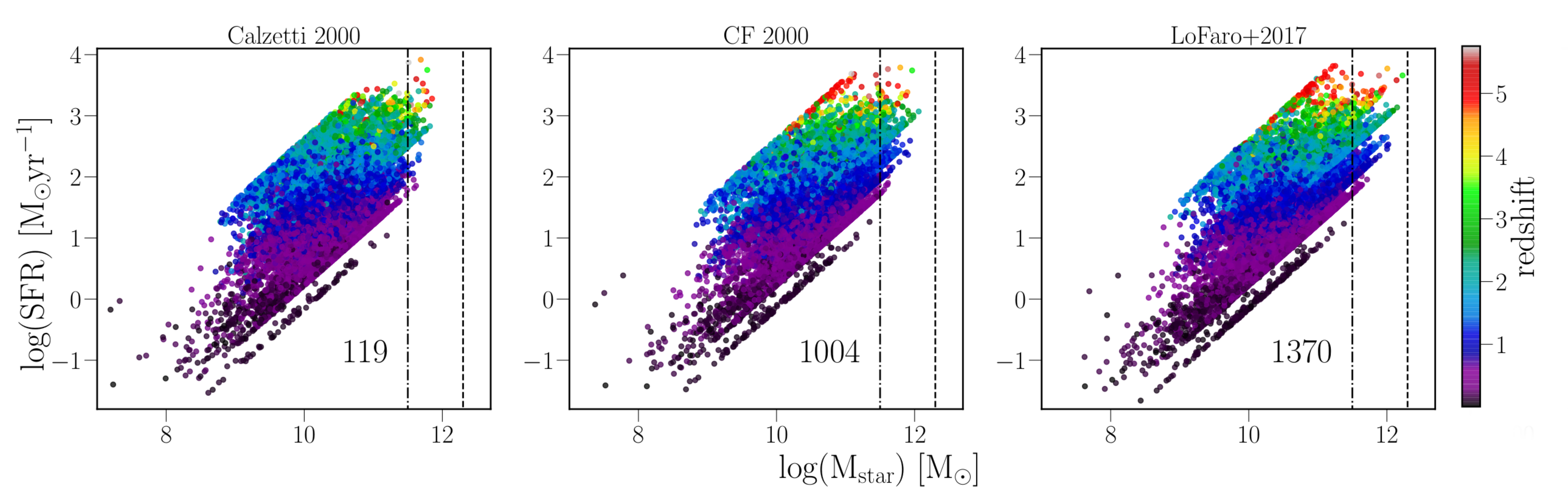}   
        \end{center}
        \caption{SFR -- $M_{star}$ relation for \cite{calzetti00}, \citetalias{CF00} and \cite{LoFaro2017} attenuation laws with colour-coded redshift values. Black dashed-dotted line represents $\rm log(M_{star})=11.5\mbox{ }[M_{\odot}]$ and dashed line $\rm log(M_{star})=12.3\mbox{ }[M_{\odot}]$. Both lines can help to distinguish the difference between the number of massive galaxies according to the scenario used for SED fitting. Numbers given in each panel correspond to the number of galaxies inside the range 11.5-12.3 $\rm log(M_{star})$.}
        \label{fig:MS_att}
\end{figure*}

The change in stellar mass has an impact on specific SFR and may also introduce a flatter shape for high-mass galaxies at the so-called main sequence galaxies (stellar mass vs. SFR relation, see Fig.~\ref{fig:MS_att}), as for the high-mass end we have more massive galaxies for \cite{CF00} and \cite{LoFaro2017} than for \cite{calzetti00} with the same SFR values. 
        
In summary the attenuation law used for SED fitting has an impact on the measure of stellar mass, but has little influence on other parameters, and it should be carefully chosen and taken into account when the properties of stellar masses are discussed. 
        
\section{Dust luminosity prediction from stellar emission}
\label{sec:Ldust_prediction}
        
The FIR emission is a key component to accurately determine the SFR of galaxies. 
The strong correlation between $\rm L_{dust}$ and SFR implies that an accurate estimate of total dust luminosity corresponds to a better SFR estimation. 
However, IR surveys are extremely expensive, and also usually  suffer from low spatial resolution, with respect to optical, and therefore result in source confusion. 
Moreover, the high-redshift galaxies are mostly explored in the visible and NIR wavelengths only.  
We investigated the relation between $\rm L_{dust}$ estimation based on different wavelength ranges. 
Based on the sample of ELAIS~N1 galaxies we verified whether or not it was possible to predict the total dust luminosity based on the optical and NIR data only. 
Two additional runs were performed. 
We ran CIGALE with the same parameters as for the initial sample of ELAIS~N1, but without IR data (rest-frame $\lambda\leq$8~$\mu$m);  we refer to that run as \texttt{stellar}. 
We also made a second run with IR data only (\textit{Herschel} PACS and SPIRE measurements), with the same models and parameters (run \texttt{IRonly}); we refer to our initial run, with all 19 bands, as \texttt{UV--IR}.   
        
For our analysis we use galaxies with at least one PACS measurement with S/N$\geq$2 and with both SPIRE~250 and SPIRE~350 measurements with S/N$\geq$2. 
Selected galaxies have at least six UV--NIR measurements and three \textit{Herschel} measurements (PACS and SPIRE)  to cover all strategic parts of the spectra for SED fitting.   
Our selection, together with $\chi^2$s criteria (Eqs.~\ref{eq:EB}), results in a sample of 586 galaxies with good coverage of the spectrum and the best photometric  measurements. 
        
\begin{figure*}[ht!]
        \begin{center}
                \subfloat[ \texttt{stellar--IR} vs  \texttt{stellar}]{\includegraphics[width=0.3\textwidth]{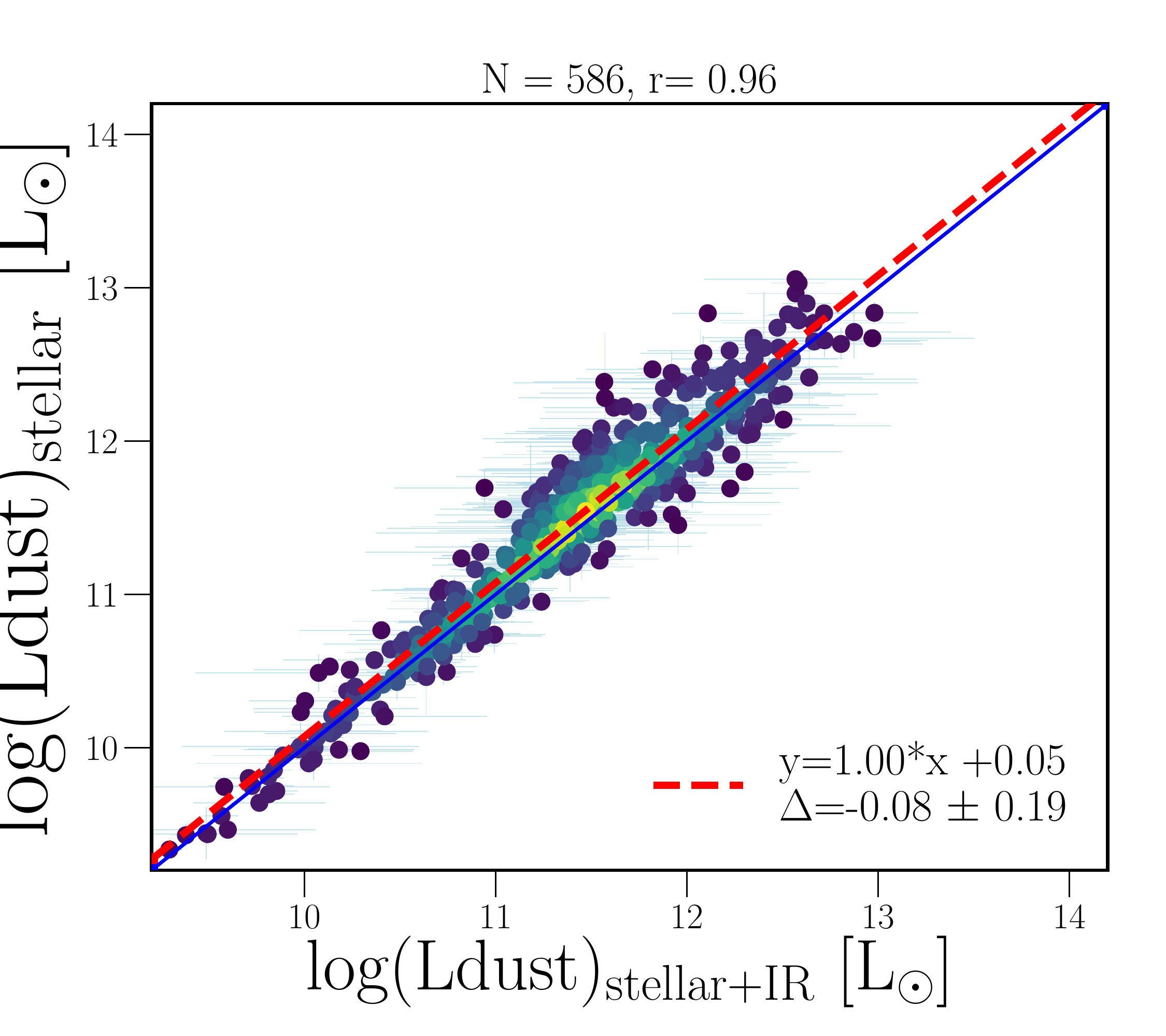}} \label{fig:Ldust_predictionA}
                \subfloat[ \texttt{stellar--IR} vs  \texttt{IRonly}]
                {\includegraphics[width=0.3\textwidth]{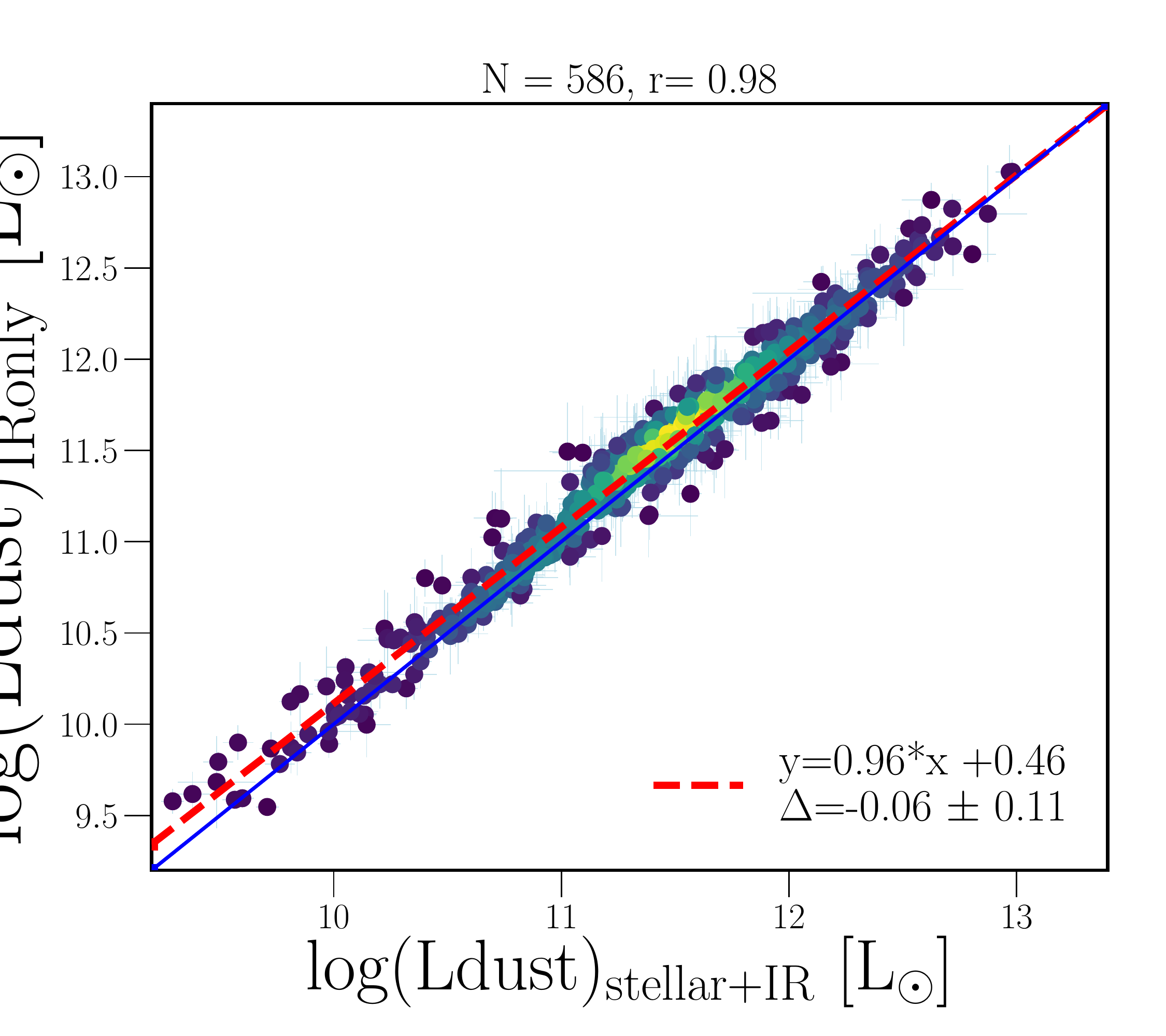}}
                \label{fig:Ldust_predictionB}
                \subfloat[ \texttt{stellar} vs  \texttt{IRonly}] {\includegraphics[width=0.3\textwidth]{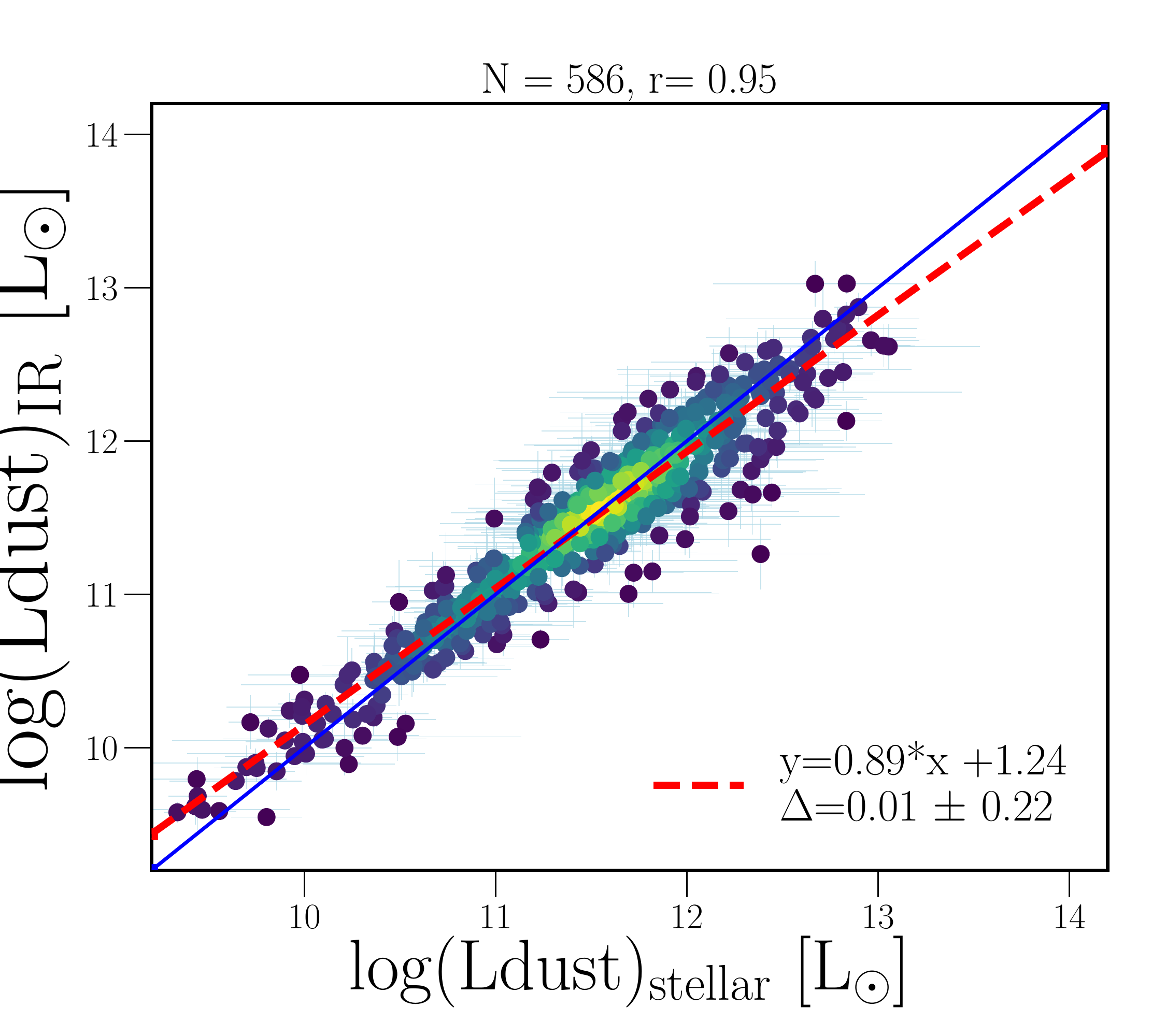}}
                \label{fig:Ldust_predictionC}
        \end{center}
        \caption{Comparison of total dust luminosity estimated using different wavelength range for 686  ELAIS~N1 galaxies with PACS~green and PACS~red measurements with S/N$\geq$1.5, SPIRE~250, and SPIRE~350 S/N$\geq$3 and at least six UV--NIR measurements; (a) results of SED fitting based on the full sets of data (x-axis) are compared with $\rm L_{dust}$ estimated based on the UV--NIR data only (y-axis); (b) results of SED fitting based on the full sets of data  (x-axis) compared to estimates from PACS and SPIRE data only (y-axis); and (c) results of SED fitting based on the UV--NIR data (x-axis) vs. estimates from PACS and SPIRE data only (y-axis).
        Blue solid lines represent 1:1 relations, while red dashed lines correspond to linear fits to the data. 
        Pearson product-moment correlation coefficient is given as an r value. 
        $\Delta$ represents the mean difference between x-axis  and y-axis  values and the standard deviation of that difference. } 
\label{fig:Ldust_prediction}
\end{figure*}
        
We check the relation between total dust luminosity for all three runs. 
First of all we find very good agreement between $\rm L_{dust}$ obtained from  \texttt{stellar--IR} and  \texttt{stellar} runs (Fig.~\ref{fig:Ldust_prediction}a). 
We find the relation to be:
        
        \begin{equation}
        \label{eq:ELAIS_N1_UVoptIR_UVop}
        \rm {log(L_{dust})_{\texttt{stellar}}=(1.00\pm0.01)\times log(L_{dust})_{\texttt{UV--IR}}+0.05}
        .\end{equation}
        
We combined a full run on (\texttt{UV--IR}) with one on \texttt{IRonly}, in which dust luminosity was calculated based on the \textit{Herschel} data only. 
Again, we found very good agreement (see Fig.~\ref{fig:Ldust_prediction}b).
        
        \begin{equation}
        \label{eq:ELAIS_N1_UVoptIR_IR}
        \rm {log(L_{dust})_{\texttt{IRonly}}=(0.96\pm 0.01)\times log(L_{dust})_{\texttt{UV--IR}}+0.46}
        .\end{equation}
        
Finally we performed the comparison of total dust luminosity calculated based on the stellar (\texttt{stellar}) data with $\rm L_{dust}$ based on the IR data only (\texttt{IRonly}) (Fig.~\ref{fig:Ldust_prediction}c). 
We have found a clear relation for $\rm L_{dust}$ values: 
\begin{equation}
\label{eq:ELAIS_N1_UVopt_IR}
\rm {log(L_{dust})_{\texttt{IRonly}}=(0.89\pm0.01)\times log(L_{dust})_{\texttt{stellar}}+1.24}
.\end{equation}
The relation has a much larger slope and scatter but again, we conclude that with CIGALE and the sets of parameters presented in Table~\ref{tab:input} we are able to predict the  dust luminosity of star forming galaxies based on the UV--NIR data only. 
The $\rm L_{dust}$ estimated based on the UV--NIR data is slightly overestimated for faint objects and underestimated for LIRGS and HLIRGS but  Eq.~\ref{eq:ELAIS_N1_UVopt_IR}, which can be used to calibrate $\rm L_{dust}$
obtained from the stellar emission only, takes this into account. 
Estimated and calibrated values of $\rm L_{dust}$  can be used for a proper estimation of SFR.
        
\section{Summary and conclusions}
\label{sec:Conclusions}

We present a strategy for  SED fitting that is applied to the \textit{Herschel} Extragalactic Legacy Project (HELP) which covers roughly 1~300~deg$^2$ of the \textit{Herschel} Space Observatory. 
We have focused on the ELAIS~N1 as a pilot field for the HELP SED fitting pipeline. 
We show how the quality of the fits and  all main physical parameters were obtained.
We present automated reliability checks to ensure that the results from the ELAIS-N1 field are of consistently high quality, with measures of the quality for every object. 
This ensures that the final HELP deliverable can be used for further statistical analysis. 
        
We introduced the two $\chi^2$s procedure to remove incorrect SED fits. 
We calculate two new $\chi^2$s,  apart from a standard  $\chi^2$ calculated in CIAGLE for the global fit: one for the stellar part ( $\rm \chi^{2}_{r,stellar}$) and one for the IR part ($\rm \chi^{2}_{r,IR}$) of the spectrum. 
We defined a threshold between $\rm \chi^{2}_{r,stellar}$ and  $\rm \chi^{2}_{r,IR}$ as 8~$\mu$m in the rest-frame. 
We demonstrate that the combination of two different $\chi^2$s can efficiently  select outliers and peculiar objects based on the energy balance assumption. 
        
We find that only {4.15}\% of galaxies of the ELAIS~N1 field show an  AGN contribution. 
We check, using four different widely known criteria for AGN selection based on the nNIR data  \citep{Stern2005AGNselection,Donley2012AGNselection,Lacy2007AGNselection,Lacy2013AGNselection}, that the majority of AGNs found based on the SED fitting procedure  (67.7\%, 58.4\%, 75.0\%, and 83.6\% respectively) fulfill the NIR criteria. 
        
We test the influence that different attenuation laws have on the main physical parameters of galaxies. 
We compare results using \cite{CF00}, \cite{calzetti00}, and \cite{LoFaro2017} recipes.  
We conclude that using different attenuation laws has almost no influence on the calculation of total dust luminosity or SFR, but we find a discrepancy between obtained stellar masses, which we find to be a direct  result of the shape of the adopted attenuation curves in NIR wavelengths. 
We  provide relations between stellar masses obtained under those three assumptions of attenuation curves. 
We find that on average the values of stellar masses for the ELAIS~N1 sample can vary up to a factor of approximately two when calculated with different attenuation laws.
We demonstrate that the recipe given by \cite{CF00} more often (for 45\% of ELAIS~N1 galaxies) outperforms that proposed by \cite{calzetti00} and \cite{LoFaro2017} (according to  $\rm \chi^2_{r,IR}$ and $\rm \chi^2_{r,stellar}$ values). 
        
We check the accuracy of estimating dust luminosity from stellar emission only and conclude that with CIGALE and the sets of parameters presented in Table~\ref{tab:input} we are able to predict  $\rm L_{dust}$ for standard IR galaxies, which preserve energy budget, based on the UV--NIR data only. 
Predicted $\rm L_{dust}$ is in very good agreement with the dust luminosity estimated based on full spectra and stellar emission only.  
We are not able to estimate monochromatic fluxes but only the total value of $\rm L_{dust}$.  
Our tests show that the SFR, tightly correlated with total dust luminosity, is also properly estimated.  
Our predictions can be used to design new surveys and as priors for the IR extraction pipeline \citep[e.g. for XID+,][]{Pearson:2018}.

%
%
\bibliographystyle{aa} 
\bibliography{KMbib} 
%

\begin{acknowledgements}
        
        The project has received funding from the European Union Seventh Framework Programme FP7/2007-2013/ under grant agreement number 60725. 
        KM has been supported by the National Science Centre (grant UMO-2013/09/D/ST9/04030).
\end{acknowledgements}

\appendix

\section{$\rm log(M_{star}) $ and SFR obtained with and without AGN module}
\label{app:with_without_test}

\begin{figure}[h!]
        \begin{center}
                \includegraphics[width=0.5\textwidth,clip]{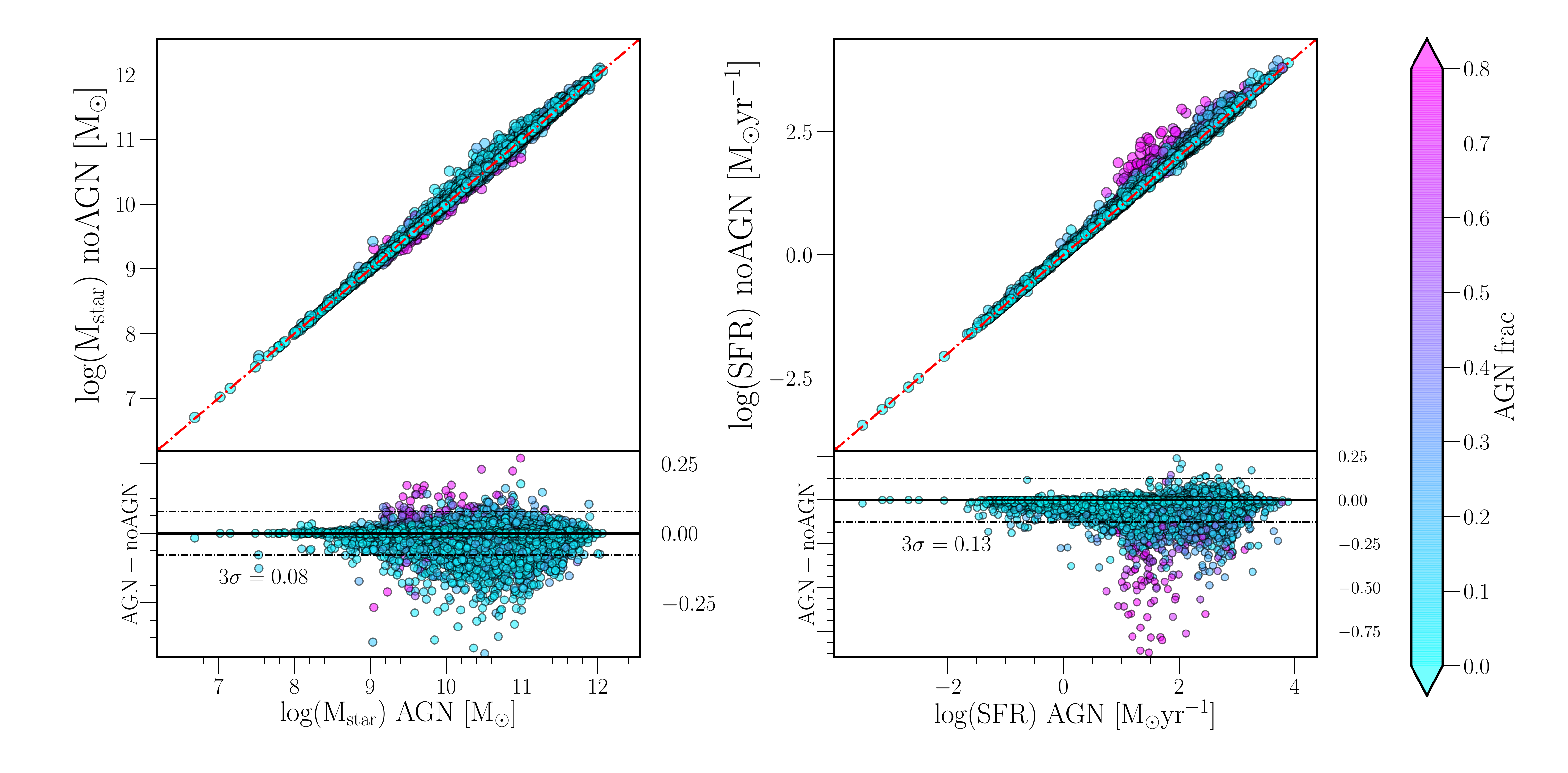}    
        \end{center}
        \caption{ Comparison of the $\rm log(M_{star})$ and SFR obtained by SED fitting without (y-axis) and with (x-axis) AGN module with colour-coded AGN fraction. Red dashed dotted lines in upper panels represent 1:1 relations. Bottom panels show difference between both runs. Black solid lines correspond to no difference, while the dashed lines represent 3$\sigma$ dispersion.}
        \label{fig:AGN_Mstar_SFR}
\end{figure}

We perform a run of the SED fitting with the same parameters for physical models but without AGN component to check how the stellar mass and the SFR differ between two runs (Fig.~\ref{fig:AGN_Mstar_SFR}).  
We find that $\rm log(M_{star})$ is generally well recovered, even for galaxies with high AGN contribution, with the standard deviation calculated for the difference of $\rm log(M_{star})$ estimated with and without AGN module equal to 0.026. 
Galaxies with significant AGN contribution have higher SFR  when calculating without the AGN component. 
For galaxies characterised by the high fraction of AGN emission the difference can be as high as 0.5 dex.
Our result is in agreement with \cite{Ciesla15} whio find that the presence of an AGN  can bias SFR estimates starting at AGN contributions higher than 10\%, reaching 100\% overestimation for AGN fraction $\sim$70\% \citep[][ Appendix~A, results for Type~2 AGNs used for our analysis]{Ciesla15}.

\section{Dust attenuation recipes for stellar emission only}
\label{app:attenuation_laws}
We check if the IR data usage can change our result presented in Sect.~\ref{sec:Dust_att} obtained with data covering wavelength range from optical to IR. 
We perform the analysis of influence of \cite{CF00} and \cite{calzetti00} attenuation laws for SED modelling of stellar emission only. 
From our test we excluded the  \cite{LoFaro2017} formula as an extreme case: this law is much flatter than those by \citetalias{CF00} and  \cite{calzetti00} and is different  suitable for very large attenuation. 

We run CIGALE with the same set of modules and parameters as for the main analysis (see Table~\ref{tab:input}) but without IR data. 
First we run it with \citetalias{CF00} dust attenuation module and, in the second step, with \cite{calzetti00}. 

We find that SED fitting with \cite{CF00} law gives better quality fits: 
71\% of galaxies were fitted better (according to the $\chi^2_r$ values) with \citetalias{CF00} than with \cite{calzetti00}. 
This results is in agreement with our previous finding presented in Sect.~\ref{sec:Dust_att}. 

Similarly to in Sect.~\ref{sec:Dust_att}, we check the influence of the dust attenuation law on estimations of stellar mass.  
For comparison we use only galaxies with at least ten optical measurements. 
In total we perform the comparison for 22~262 galaxies.

\begin{figure*}[]
        \begin{center}
                \subfloat[$\rm log(L_{dust})$]
                {\includegraphics[width=0.31\textwidth]{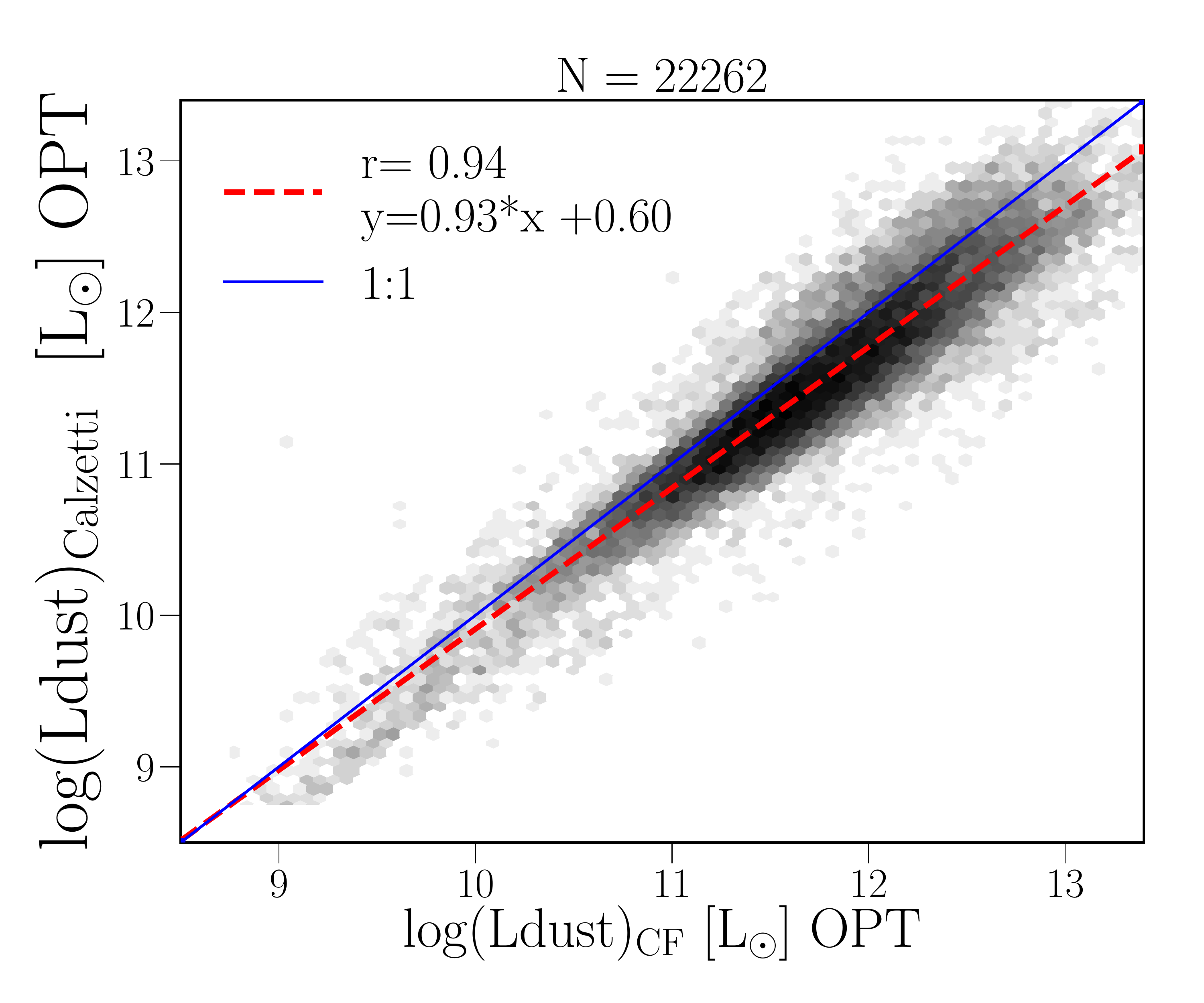}} \label{fig:stellar_Ldust}
                \subfloat[$\rm log(SFR)$]
                {\includegraphics[width=0.31\textwidth]{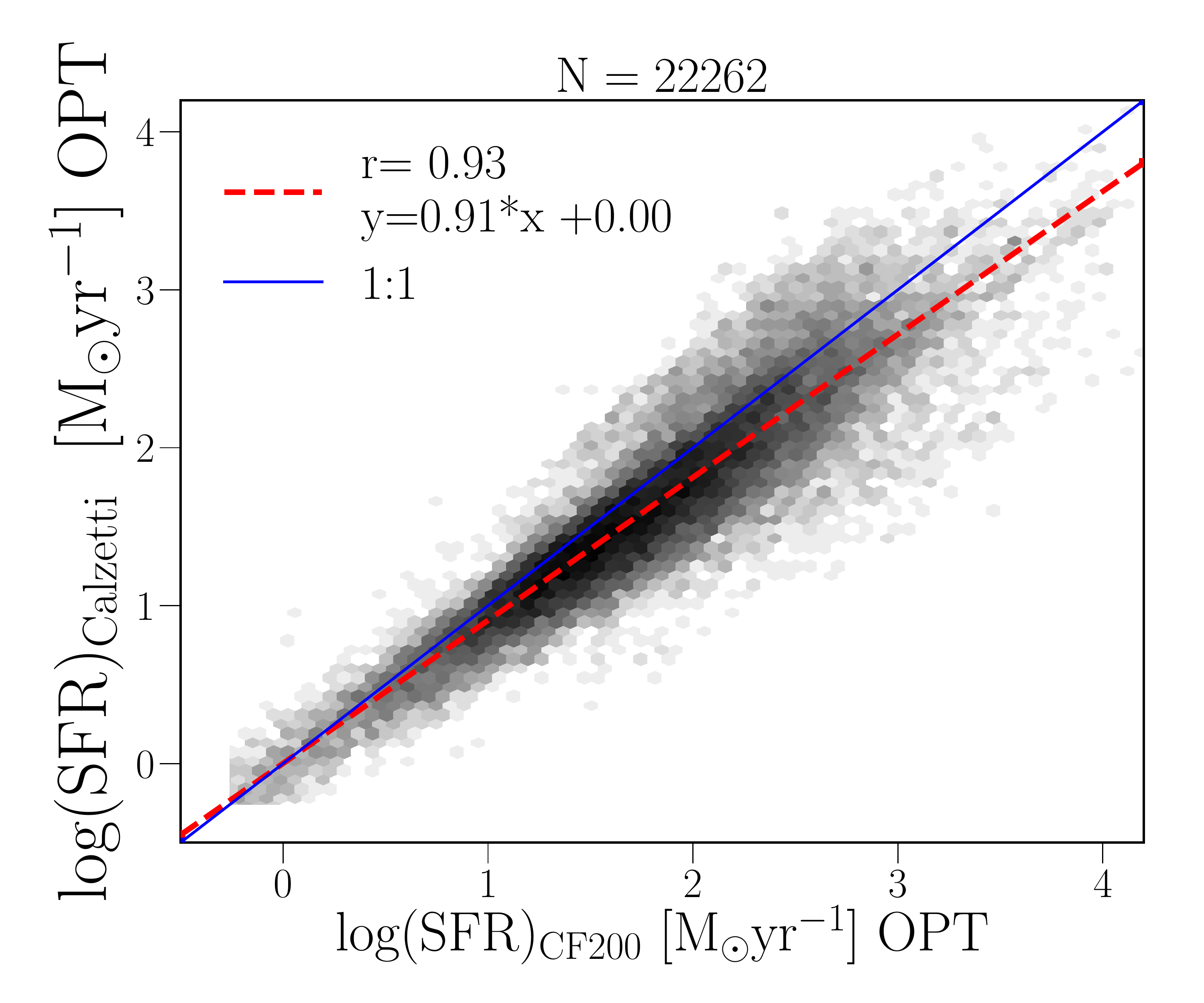}}\label{fig:stellar_SFR}
                \subfloat[ [$\rm log(M_{star})$ ] {\includegraphics[width=0.31\textwidth]{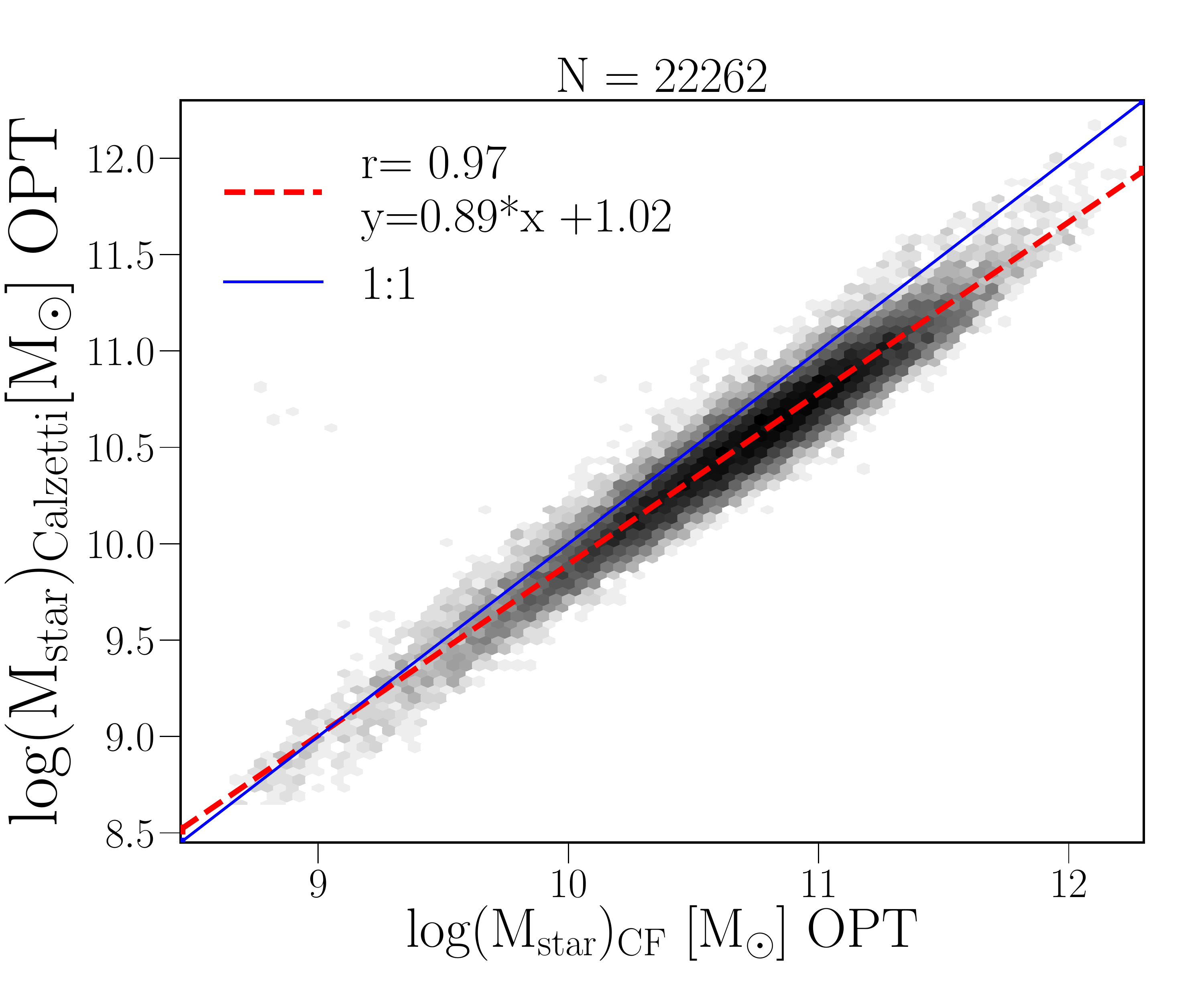}}\label{fig:stellar_Mstar}
        \end{center}
        \caption{Comparison of estimated $\rm log(L_{dust})$ (\textit{panel a}), $\rm log(SFR)$ (\textit{panel b}) and $\rm log(M_{star})$ (\textit{panel c}) for a sample of 22~262 galaxies fitted with \citealp{calzetti00} (y-axis) and \citealp{CF00} (x-axis) laws. Blue solid line represents 1:1 relation, while fitted relation is marked as a red dashed line. $\texttt{r}$ represents calculated Pearson product-moment correlation coefficient.} 
        \label{fig:stellar_emission}
\end{figure*}

The $\rm L_{dust}$ (Fig.~\ref{fig:stellar_emission}~a),  and consistently the SFR (Fig.~\ref{fig:stellar_emission}~b), obtained from two runs are scattered in the region of ULIRGs and HLIRGs (we obtained a similar scatter in Fig.~\ref{fig:Ldust_prediction}~c comparing the the total dust luminosity estimated using different wavelength ranges. 
As $\rm L_{dust}$ and SFR are tightly related, the same effect is visible in Fig.~\ref{fig:stellar_emission}~b.
Figure~\ref{fig:stellar_emission}~c shows the relation between $\rm log(M_{star})$ obtained with the attenuation laws of  \cite{CF00} and \cite{calzetti00} by fitting models for the stellar part only. 
Obtained relation (Eq.~\ref{eq:Mstar_stellar_emission}) is in a perfect agreement with Eq.~\ref{eq:Mstar_full_SED_fitting} obtained for the full spectral fitting (including IR data).
\begin{equation}
\label{eq:Mstar_stellar_emission}
log(M_{star})_{Calzetti} = 0.97 \times log(M_{star})_{CF} + 1.02 
.\end{equation}

We conclude that the relation between stellar mass estimated with attenution laws by \citetalias{CF00} and \cite{calzetti00} is valid with and without IR data included for SED fitting.

\begin{figure*}[]
        \centering
        \includegraphics[width=0.85\textwidth, clip]{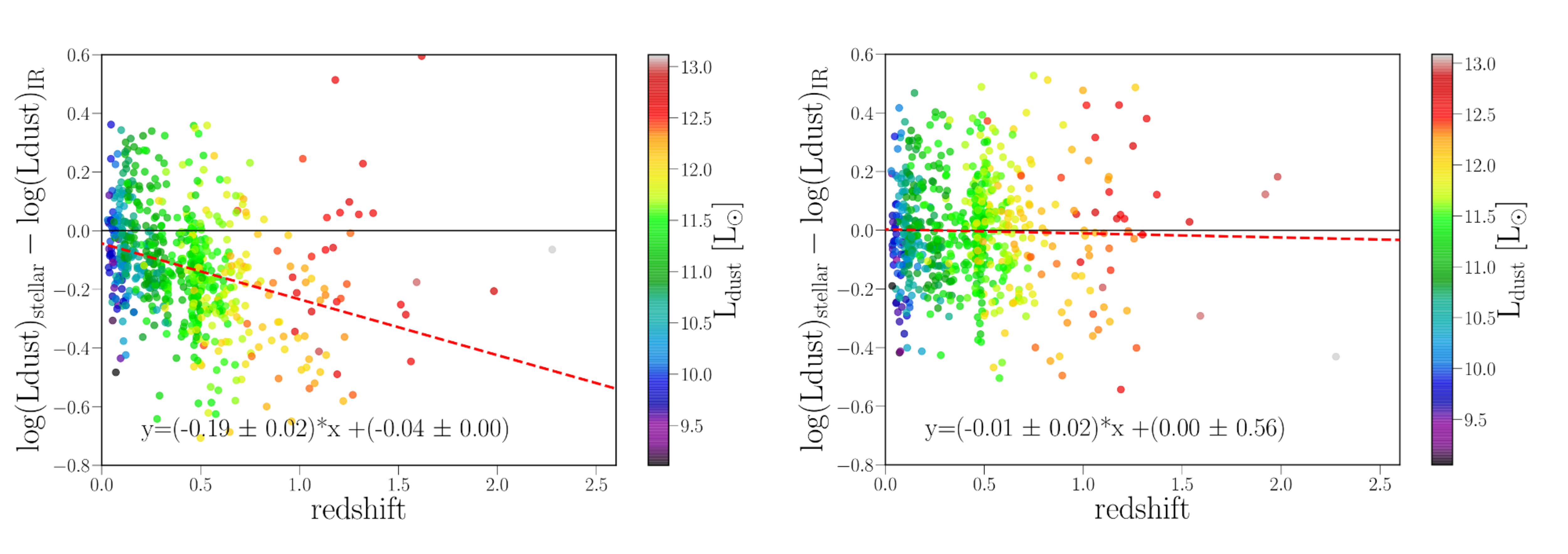}
        \caption{The difference between $\rm L_{dust}$ calculated from the stellar emission and IR emission only in function of $\rm L_{dust}$ estimated based on the stellar + IR emission. The left column represents a run with the \cite{calzetti00} law while the right column corresponds to the run with the \cite{CF00} recipe.  } 
        \label{fig:app_comparison_Ldust}
\end{figure*}

We check whether or not using \citetalias{CF00} or \cite{calzetti00} attenuation law to calculate $\rm L_{dust}$ from the stellar emission only can give some bias in function of redshift or $\rm L_{dust}$ estimated based on the full (optical + IR emission) SED fitting. 
We find that in both cases the scatter is very similar but in case of the \cite{calzetti00} recipe we find a clear dependence with $\rm L_{dust}$ and redshift: for bright IR galaxies, the $\rm L_{dust}$ calculated based on the stellar emission only is underestimated. 
The same relation applies for more distant objects (Fig.~\ref{fig:app_comparison_Ldust}).

\end{document}